\documentclass[useAMS,usenatbib]{mn2e}
\usepackage{graphicx}
\usepackage{xcolor}
\usepackage{amssymb,amsmath,bm}
\usepackage{hyperref}
\hypersetup{draft}
\usepackage[export]{adjustbox}
\usepackage{multirow}

\hypersetup{
    colorlinks=false,
    citecolor=blue,
    linkcolor=blue,
    linktoc=page
}

\newcommand{\aap}{Astron.\ Astrophys.\ }
\newcommand{\apj}{Astrophys.\ J.\  }
\newcommand{\apjl}{Astrophys.\ J.\ Lett.\  }
\newcommand{\apjs}{Astrophys.\ J.\ Suppl.\  }
\newcommand{\mnras}{Mon.\ Not.\ R.\ Astron.\ Soc.\ }

\newcommand{\physrep}{{\em Phys.\ Rep.}}
\newcommand{\pasj}{Publ.\ Astron.\ Soc.\ Japan\ }
\newcommand{\prd}{Phys.\ Rev.\ D\ }
\newcommand{\jcap}{Journal of Cosmology and Astro-Particle Physics}

\newcommand{\avg}[1]{\ensuremath{\left\langle \,#1\, \right\rangle}}

\newcommand{\der}{\ensuremath{{\rm d}}}

\newcommand{\tj}[6]{ \begin{pmatrix}
   #1 & #2 & #3 \\
   #4 & #5 & #6 
  \end{pmatrix}}

\newcommand{\Gj}[6]{ \begin{Bmatrix}
   #1 & #2 & #3 \\
   #4 & #5 & #6 
  \end{Bmatrix}}
  
\newcommand{\be}{\begin{equation}}
\newcommand{\ee}{\end{equation}}

\newcommand{\bea}{\begin{eqnarray}}
\newcommand{\eea}{\end{eqnarray}}
\newcommand{\bdm}{\begin{displaymath}}
\newcommand{\edm}{\end{displaymath}}

\def\Mpc{\, h^{-1} \, {\rm Mpc}}

\def\Gpc{\, h^{-1} \, {\rm Gpc}}

\def\kMpc{\, h \, {\rm Mpc}^{-1}}

\def\dk{\frac{\mathrm{d}^3\,k}{(2\pi)^3}\,}

\newcommand{\eq}[1]{Eq.~(\ref{#1})}
\newcommand{\eqs}[2]{Eqs.~(\ref{#1},\ref{#2})}
\newcommand{\fig}[1]{Figure~\ref{#1}}

\newcommand{\vb}[1]{\mathbf{#1}}
\def\ie{{\em i.e.}~}
\def\eg{{{\em e.g.}~}}

\title[Wide angle effects]{Beyond the plane-parallel approximation for redshift surveys}

\author[E. Castorina, M. White]
{Emanuele Castorina$^{1,2}$\thanks{e-mail: ecastorina@berkeley.edu}, 
Martin White$^{1,2}$\thanks{e-mail: mwhite@berkeley.edu}
 \\~\\
\footnotesize
\footnotesize
$^1$Berkeley Center for Cosmological Physics, University of California, Berkeley, CA 94720, USA\\
$^2$Lawrence Berkeley National Laboratory, 1 Cyclotron Road, Berkeley, CA 93720, USA\\
}

\begin{document}
\maketitle 

\begin{abstract}
Redshift space distortions privilege the location of the observer in cosmological redshift surveys, breaking the translational symmetry of the underlying theory.  This violation of statistical homogeneity has consequences for the modeling of clustering observables, leading to what are frequently called `wide angle effects'.  We study these effects analytically, computing their signature in the clustering of the multipoles in configuration and Fourier space. We take into account both physical wide angle contributions as well as the terms generated by the galaxy selection function. Similar considerations also affect the way power spectrum estimators are constructed. We quantify in an analytical way the biases which enter and clarify the relation between what we measure and the underlying theoretical modeling. The presence of an angular window function is also discussed. Motivated by this analysis we present new estimators for the three dimensional Cartesian power spectrum and bispectrum multipoles written in terms of spherical Fourier-Bessel coefficients. We show how the latter have several interesting properties, allowing in particular a clear separation between angular and radial modes. 
\end{abstract}


\section{Introduction}
\label{sec:intro}
The interpretation of clustering measurements in spectroscopic surveys is complicated by the fact we do not observe the true distance to any object, as its measured redshift receives a contribution from the line-of-sight (LOS) component of its peculiar velocity \citep{Kai87,H92,H98,Pea99}.  This effect, known as redshift space distortions (RSD), breaks the statistical isotropy of our theories and makes cosmological analyses more complex.  At the same time, the signal embedded in RSD offers new ways of extracting information from a dataset, enabling us for instance to measure the rate of growth of large-scale structure and to better probe neutrino masses and theories of modified gravity \citep{Lesgourgues,Wei13}. The study of RSD remains a major area of theoretical research in cosmology.

\begin{figure}
\begin{center}
\begin{picture}(250,250)
\Large
\thicklines
\qbezier(125.0,0.0)(125.0,112.5)(125.0,218.8)
\qbezier(125.0,0.0)(156.2,125.0)(187.5,250.0)
\qbezier(125.0,0.0)(93.8,93.8)(62.5,187.5)
\qbezier(62.5,187.5)(125.0,218.8)(187.5,250.0)
\thinlines
\multiput(125,219)(0,6){5}{\line(0,1){3}}
\put(110,60){$\displaystyle{\frac{\theta}{2}}$}
\put(128,60){$\displaystyle{\frac{\theta}{2}}$}
\put( 70,120){$\vec{s}_1$}
\put(160,120){$\vec{s}_2$}
\put(130,180){$\vec{d}$}
\put( 70,182){\rotatebox{27}{$s(1-t)$}}
\put(160,227){\rotatebox{27}{$st$}}
\put(115,205){$\phi$}
\put(160,250){\vector(-2,-1){75}}
\put(110,230){$\vec{s}$}
\end{picture}
\caption{The assumed geometry and angles.  The two galaxies lie at $\vec{s}_1$
and $\vec{s}_2$, with separation vector $\vec{s}=\vec{s}_1-\vec{s}_2$ and
enclosed angle $\theta$.
We take the line of sight to be parallel to the angle bisector, $\vec{d}$,
which divides $\vec{s}$ into parts of lengths $st$ and $s(1-t)$.  The
separation vector, $\vec{s}$, makes an angle $\phi$ with the line of sight
direction, $\hat{d}$. \label{fig:conf}}
\end{center}
\label{fig:triangle}
\end{figure}
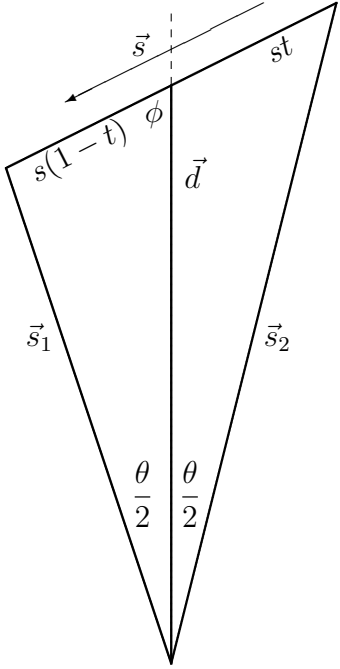

The study of RSD within cosmological perturbation theory was pioneered by \citet{Kai87}, who showed that at linear order peculiar velocities introduce a distinct signature in the clustering of galaxies, squashing the two-point correlation function perpendicularly to the LOS.  Defining $\vb{r}$ to be the true coordinate of a galaxy, the observed redshift space coordinate is 
\begin{equation}
  \vb{s} \equiv \vb{r} +f \hat{r}\, \vb{u}\cdot\hat{r}
\end{equation}
where $f$ is the linear growth factor and $\vb{u}$ is related to the the peculiar velocity $\vb{v}$ and the Hubble parameter $H$, $\vb{v} =  aH f \vb{u}$. In linear theory we can write the following relation between real space and the redshift space field \citep{Kai87,H92,H98}.
\begin{equation}
\label{eq:RSD}
\delta_s(\vb{r}) = \left[1+ f\left(\frac{\partial^2}{\partial r^2}+\frac{\alpha(\vb{r})\partial}{r \partial r}\right)\nabla^{-2} \right]\delta(\vb{r})\equiv \mathcal{O}_s\delta(\vb{r})\;,
\end{equation}
where $\alpha(\vb{r})$ is the logarithmic derivative of the galaxy selection function
\begin{equation}
\alpha(\vb{r}) \equiv \frac{\der \ln r^2 \bar{n}(\vb{r})}{\der \ln r}
\end{equation}
which we will assume varies slowly with $r$.
Being the product of two operators in configuration space, the redshift space density in Fourier space will be a convolution \citep{ZH96}
\begin{equation}
\delta_s(\vb{k}) = \int \dk \mathcal{O}(\vb{k},\vb{k}') \delta(\vb{k}')
\quad ,
\end{equation}
and in fact $\mathcal{O}(\vb{k},\vb{k}')$ has a strong $k$-dependence.
It is clear by looking at the operators defined above that, as a result of having chosen a preferential observer, RSD partially break statistical homogeneity and isotropy of $n$-point functions. The only symmetry one is left with is rotational symmetry around the observer and azimuthal symmetry about the line of sight.
Loss of translation invariance means the power spectrum is not diagonal
\begin{equation}
\label{eq:RSDk}
  \avg{\delta_s(\vb{k}_1)\delta(\vb{k}_2)}=P_{s}(\vb{k}_1,\vb{k}_2) 
\end{equation}
and the configuration space the 2-point correlation function will no longer only depend on the relative separation between two galaxies but on the triangle formed by the observer and the two galaxies\footnote{Another way to think of the mode coupling is that it arises from aliasing due to the projection of the spherically symmetric configuration selected by the physics onto the planar (triangle) configuration of the observation.  See \citet{Sza98}.}.  Fig.~\ref{fig:triangle} shows the geometry of the problem.  Parameterizing this triangle by the pair separation, $s\equiv|s_1-s_2|$, the distance to the pair, $d$, and the cosine $\mu\equiv \hat{s}\cdot \hat{d}$:
\begin{equation}
  \avg{\delta_s(\vb{s}_1)\delta(\vb{s}_2)}=\xi_{s}(\vb{s}_1,\vb{s}_2) = \xi_{s}(s,d,\mu)
\end{equation}
The familiar expression in \cite{Kai87} be recovered in the plane parallel approximation, \ie all LOS's are parallel, for instance along the $z$-direction: $\hat{s}_1\simeq\hat{s}_2\simeq\hat{d}\simeq\hat{z}$. This is equivalent to assume the sky is flat.
In this case, defining $s\equiv|s_1-s_2|$ and $\mu \equiv \hat{s}\cdot \hat{z}$, one has
\begin{equation}
  \avg{\delta_s(\vb{s}_1)\delta(\vb{s}_2)}\simeq \xi_{s}^{pp}(s,\mu)
\end{equation}
where the superscript $pp$ indicates the assumption of the plane-parallel approximation.
It is conventional to expand the $\mu$ dependence in Legendre polynomials, $\mathcal{L}_\ell$, via
\begin{equation}
\label{eq:Kaiserx}
  \xi_{s}^{pp}(s,\mu)  = \sum_\ell \xi_\ell^{pp}(s)\mathcal{L}_\ell(\mu)
\end{equation}
Similarly for the power spectrum in the plane parallel limit
\begin{equation}
  \avg{\delta_s(\vb{k}_1)\delta(\vb{k}_2)} \simeq
    (2\pi)^3 \delta_D^{(3)}(\vb{k}_1+\vb{k}_2) P_{s}^{pp}(k,\mu_k)
\label{eqn:plane_parallel_P}
\end{equation}
and we can expand
\begin{equation}
\label{eq:Kaiserk}
  P_{s}^{pp}(k,\mu_k) = \sum_\ell P_\ell^{pp}(k)\mathcal{L}_\ell(\mu_k) 
\end{equation}
where $\mu_k \equiv \hat{k}\cdot \hat{z}$ and in linear theory
\begin{align}
\label{eq:PkK}
P_{0}^{pp}(k) &= P(k) \left(1+\frac{2}{3}f + \frac{1}{5}f^2 \right)\notag \\
P_{2}^{pp}(k) &= P(k)\left(\frac{4}{3}f + \frac{4}{7}f^2\right) \notag \\
P_{4}^{pp}(k) &= P(k)\left(\frac{8}{35}f^2 \right) \quad .
\end{align}

In configuration space, the same limit in \eq{eq:Kaiserx} can be obtained from the general case assuming the galaxy separations one is interested in are much smaller than the distance between the observer and the galaxies, i.e.~$s/d\to 0$ \citep{Kai87,H92,HC96}.
However, as shown for the first time by \citet{ZH96}, there is no well defined approximation in Fourier space that would lead to \eq{eq:Kaiserk} from \eq{eq:RSDk}. This is mostly a consequence of the fact that scalar products between vectors in configuration space and Fourier space we used to define $\mu_k$ are ill-defined in Fourier space.

This result poses important questions for the interpretation of Fourier space analysis of galaxy surveys, which are now observing large fractions of the sky and are no longer in the small angle/plane parallel limit. This affects both the estimators of the power spectrum multipoles and the analytical predictions, as we want to use a model that is a close as possible to what we actually measure. In particular, most if not all perturbation theory models for the power spectrum beyond linear theory have been written down in the plane parallel-approximation, as it vastly simplifies the calculations
\citep[e.g.][for recent examples]{TNS10,ReiWhi11,CLPT,Oku15,Whi15,VCW16,Perko16}.

The scope of this paper is to present a self-contained analytical calculation of wide angle affects in galaxy surveys. 
We are certainly not the first to try to address this issue. 
From the early work of \citet{H92,HC96,H98,ZH96,Sza98}, to more recent works by \citet{Sza04,Dat07,PapSza08,ShaLew08,Bonvin,Raccanelli,YS15,Slepian15,Rei16} there is a large literature on this topic, and our analysis relies heavily on it.
However what is still missing is a analytical model able to capture all the RSD terms entering the expression for the multipoles, at least in linear theory, and their relation to the estimators currently used in the analysis of galaxy surveys. 

This paper is organized as follows. In Section \ref{sec:linear_theory} we set up some more notation and discuss the plane parallel limit. 
We compute leading order correction to the plane-parallel formulae in both configuration space and Fourier space.
Then in Section \ref{sec:fourier_estimators} we study commonly used estimators of the power spectrum multipoles and their relation to analytical models. We quantify the error that the Fast Fourier Transform (FFT) estimators presented in \citet{Bia15,Sco15,Han17,Wil17} make compared to the true underlying power spectrum as a result of a small angle approximation.
In Section \ref{sec:sFB} we describe an alternative basis which respects the symmetries of the problem: the spherical Fourier-Bessel (sFB) basis.  We present new estimators of the power spectrum based on sFB coefficients that we believe are well suited for analysis of data on the curved sky.  We also make connection to other 2-point statistics and the plane-parallel limit. Finite volume and masking effects are also discussed.
We then draw our conclusions and discuss future directions in Section \ref{sec:conclusions}.
We defer some technical details to a series of Appendices, which discuss aspects of geometry, useful mathematical identities, a recap of the main linear theory results, seperable power spectrum estimators to $\mathcal{O}(\theta^2)$ and an estimator for the bispectrum in the Fourier-Bessel basis.

\begin{figure*}
\centering
\resizebox{\textwidth}{!}{\includegraphics{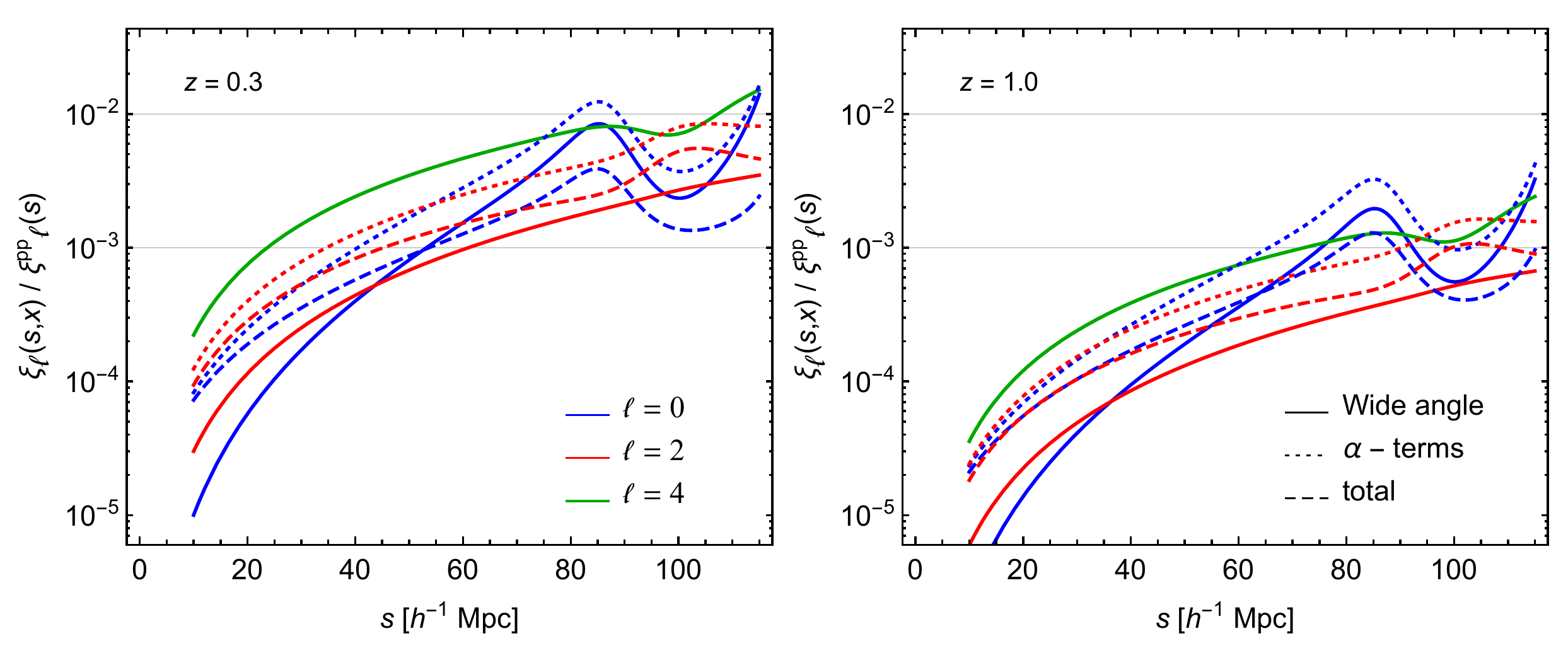}}
\caption{The ratio between multipoles of the correlation function in the plane-parallel limit and the leading order wide angle contribution. Physical wide-angle terms are depicted with the continuous lines, selection function terms with dotted ones, and the sum of the two with dashed lines. Left panel shows the ratio at $z=0.3$, whereas the right panel at $z=1$.}
\label{fig:xiL}
\end{figure*}

\section{Wide angle corrections in linear theory}
\label{sec:linear_theory}

\subsection{Configuration space}

The prototypical redshift space configuration is depicted in \fig{fig:triangle}.
An observer $O$ is looking at two galaxies at $\vb{s}_1$ and $\vb{s}_2$. The pair separation is $\vb{s} = \vb{s}_1 - \vb{s}_2$ and the LOS $\vb{d}$ is defined as the bisector of the angle $\theta$ between $\vb{s}_1$ and $\vb{s}_2$. (Another possible definition of the bisector is the midpoint of $\vb{s}$, which corresponds to the angle bisector for isosceles triangles but is otherwise different.  For our purposes the former proved to be the most convenient. Appendix \ref{app:geometry} provides the mapping between the two possible choice of LOS.)
The angle between the LOS and the separation vector is $\phi$, and $\cos(\phi)\equiv \mu$. 
The parameter $t\in [0,1]$ describes how the LOS intercepts $s$,
\begin{align}
\label{eq:bis}
\vb{s}_1 &= \vb{d} + (1-t)\, \vb{s} \notag \\
\vb{s}_2 &= \vb{d} - t \,\vb{s} 
\end{align}
Any triangle is specified by three numbers, either two lengths and one angle or two angles and one length.  If we choose the latter and if the LOS is defined by the bisector, the correlation function can be decomposed as a double Legendre series,
\begin{equation}
\label{eq:xiLL}
\xi_s(s,\mu,\theta) = \sum_{\ell_1 \ell_2} C_{\ell_1 \ell_2}(s) \mathcal{L}_{\ell_1}\left(\cos\frac{\theta}{2}\right) \mathcal{L}_{\ell_2}(\mu)
\end{equation}
with the coefficients $C_{\ell_1 \ell_2}(s)$ predicted by perturbation theory (e.g.~\citealt{Sza98,Sza04}; Appendix \ref{app:derivation_sketch}).  From the properties of $\mathcal{L}_\ell$ it is clear that as $\theta\rightarrow0$ \eq{eq:xiLL} reduces to \eq{eq:Kaiserx}, and \citet{Sza98,Sza04} show that the linear theory results reduce to those derived in \cite{Kai87,H92}.  Whereas the first term in parenthesis in \eq{eq:RSD} always contributes only a finite number of coefficients, the selection function term generates an infinite number of them.

We are interested in the leading order corrections to \eq{eq:Kaiserx}.  As a small parameter we use $x\equiv s/d$, rather than $\theta$, as this will prove more convenient later and will allow more efficient computation of the power spectrum multipoles \citep[see also][]{Rei16}. 
We leave the details of the calculation to Appendix A and present here the main results. Writing \eq{eq:xiLL} as a Taylor series in $x$ boils down to expanding $\mathcal{L}_\ell(\cos\theta/2)$ in powers of $x$,
\begin{equation}
  \mathcal{L}_\ell\left(\cos\frac{\theta}{2}\right) \simeq
  1 - x^2\frac{\ell(\ell+1)}{16}\left[1-\mu^2\right] + \cdots
\end{equation}
which shows the first important result that wide angle corrections start at $\mathcal{O}(x^2)$. For a pair of galaxies separated by the BAO scale at redshift $z=1$ we find $x\simeq 0.045$. 
We want to recast the expression for the correlation function into the following form\footnote{Note that we will use $\xi_\ell$ without a superscript and with two arguments to refer to this wide-angle quantity, while $\xi_\ell^{(n)}$ with the superscript -- defined in Eq.~(\ref{eqn:xi_ell_n}) -- is an integral of the linear theory power spectrum.}
\begin{equation}
\label{eq:xiell}
\xi_s(s,d,\mu) = \sum_{\ell} \xi_\ell(s,d) \mathcal{L}_{\ell}(\mu) 
\end{equation} 
where the $\xi_\ell(s,d)$'s are series expansions in the wide angle parameter $x$, 
\begin{equation}
\label{eq:xiO}
\xi_\ell(s,d) =a_\ell^{(0)}(s)\, x^0+ a_\ell^{(2)}(s) \,x^2+...
\end{equation}
The $a_\ell^{(0)}$'s are nothing else than the plane parallel terms given in \eq{eq:Kaiserx}. 
In linear theory these terms are
\begin{align}
\label{eq:xiK}
\xi_{0}^{pp}(s) &= \xi_0^{(0)}(s) \left(1+\frac{2}{3}f + \frac{1}{5}f^2 \right)\notag \\
\xi_{2}^{pp}(s) &= \xi_2^{(0)}(s)\left(-\frac{4}{3}f - \frac{4}{7}f^2\right) \notag \\
\xi_{4}^{pp}(s) &= \xi_4^{(0)}(s)\left(\frac{8}{35}f^2 \right)
\end{align}
where\footnote{Beware: one often finds similar definitions with an additional $i^\ell$.  We do not include this factor.}
\begin{equation}
  \xi_\ell^{(n)}(s,d) = \int\frac{k^2\,dk}{2\pi^2}
                        \ (kd)^{-n} P(k)j_\ell(ks)  \quad .
\label{eqn:xi_ell_n}                      
\end{equation}
Note that for power-law $P(k)$, $\xi_\ell^{(n)}\sim x^n\xi_\ell^{(0)}$ and we shall use this scaling below.

The double-derivative piece in \eq{eq:RSD} generates the following wide-angle contributions 
\begin{align}
\label{eq:xiWA}
  \xi_0(s,d) &\ni -\frac{4f^2}{45}x^2\ \xi_0^{(0)}(s) - \frac{f(9+f)}{45}x^2\ \xi_2^{(0)}(s)  \notag \\
  \xi_2(s,d) &\ni  \frac{4f^2}{45}x^2\ \xi_0^{(0)}(s) + \frac{f(189+53f)}{441}x^2\ \xi_2^{(0)}(s)
  \notag \\
  &            -\frac{4f^2}{245}x^2\ \xi_4^{(0)}(s) \notag \\
  \xi_4(s,d) &\ni -\frac{8f(7+3f)}{245}x^2\ \xi_2^{(0)}(s) + \frac{4f^2}{245}x^2\ \xi_4^{(0)}(s)
\end{align}

To estimate the correction due to the selection function we have to make a choice for $\bar{n}(\vb{s})$. In the case of a uniform sample $\alpha(s) = 2$ and we obtain the following new terms
\begin{align}
\label{eq:xialpha}
\xi_0(s,d) &\ni \frac{4f^2}{3}\ \xi_0^{(2)}
  + \frac{2}{3}f(1-f)x\ \xi_1^{(1)} \notag \\
  \xi_2(s,d) &\ni -\frac{8}{3}f^2\ \xi_2^{(2)}
     -\frac{8}{15}f(5+f)x\ \xi_1^{(1)} +\frac{4}{5}f^2x\ \xi_2^{(1)}
\end{align}
and no contribution to the hexadecapole. 

\fig{fig:xiL} shows the ratio of the wide angle terms to the plane-parallel multipoles for an observation at $z=0.3$ (left panel), with $d(z=0.3) = 1546 \Mpc$, and  $z=1$ (right panel), with $d(z=1) = 2367 \Mpc$. 
The first thing worth noticing is that for uniform samples, the physical term and the geometric term have opposite sign and partially cancel each other. We stress that this is an accidental cancellation, and it will be different for more generic and realistic selection functions. As expected, going to higher redshift makes wide angle corrections in the selection function less and less important, with the highest multipole presenting the largest difference. Our analysis indicates that in the modeling of the full shape of the correlation wide angle terms terms can be safely neglected if the required accuracy is 1\%.

\begin{figure}
\centering
\resizebox{\columnwidth}{!}{\includegraphics{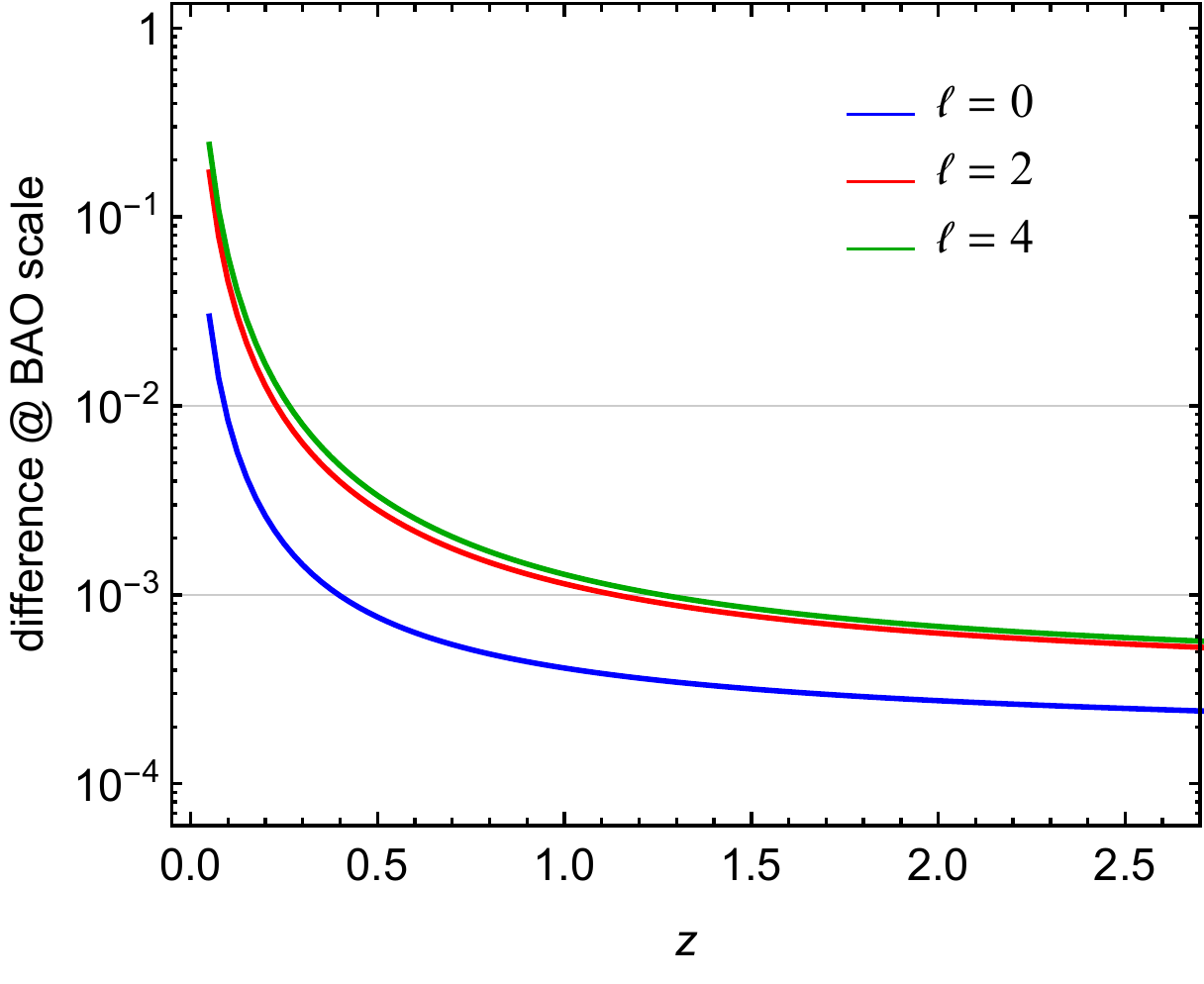}}
\caption{Contribution at the BAO scale, $s\simeq105 \Mpc$, to a configuration space multipole $\ell$ from wide angle terms with $\ell'\ne \ell$. These terms are out of phase, \ie have different BAO position, than the $\ell^{\rm th}$ multipole.}\label{fig:BAO}
\end{figure}

Next we investigate the effect of wide angles on BAO scales. The way BAO information is usually extracted from data requires a template and a marginalization over the broadband shape of the power spectrum/correlation function, see for instance the most recent results of the BOSS survey \citep{Beu17,Ross17}. This means broadband components tend not to cause shifts in the inferred distances. On the other hand terms beyond the PP approximation mix different $\xi_\ell^{(n)}$, see for instance \eqs{eq:xiWA}{eq:xialpha}, which are out of phase from each other and this could potentially move the position of the BAO peak in the full correlation function.

To isolate this effect, \fig{fig:BAO} shows, as a function of redshift, the contribution to each multipole $\ell$ at the BAO scale of wide angle terms with $\ell'\ne\ell$.
At low enough redshift, we find contributions at the percent level or even larger. This difference doesn't automatically translate into a bias in the distance estimates, but it suggests that some care is warranted.  Luckily the effect is isolated to large scales where linear theory is adequate, and thus the impact of wide-angle effects can be included in the template fitting, which would ameliorate any potential for biases to enter.  We shall return to this idea later.

\subsection{Fourier space}
\label{sec:fourier_space}

\begin{figure*}
\centering
\resizebox{\textwidth}{!}{\includegraphics{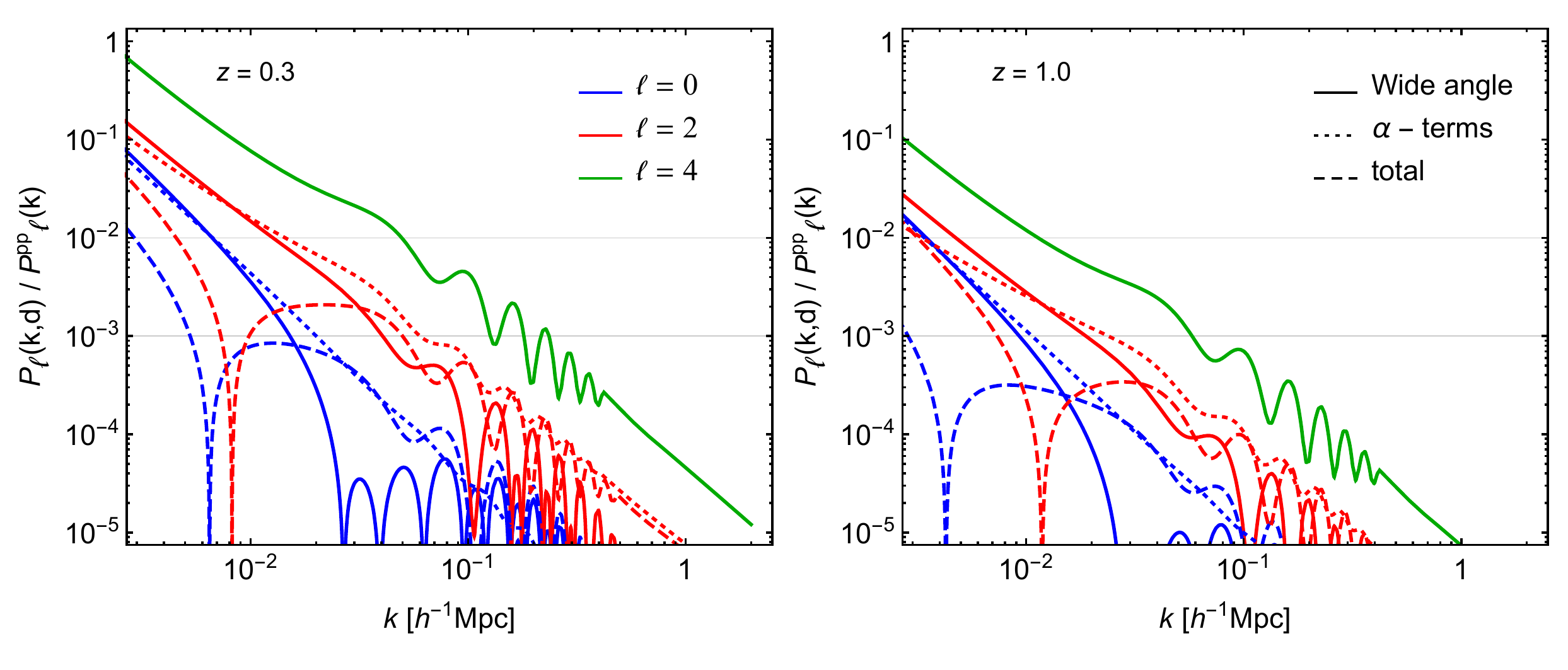}}
\caption{Similar to \fig{fig:xiL} but now for the multipoles in Fourier space. Note the cancellation between wide angle and selection functions terms in both the quadrupoles and the hexadecapole.}
\label{fig:PkL}
\end{figure*}
As emphasized by \citet{ZH96}, upon dropping the plane-parallel approximation, we must be careful in defining the power spectrum (and its multipoles).
A convenient form for our purposes is the ``local'', or LOS-dependent, power spectrum, which is defined in a mixed space \citep{Sco15,Rei16}.  Specifically
\begin{equation}
  P(\vb{k},\vb{d}) \equiv \int \mathrm{d}^3 s
    \, \xi(\vb{s},\vb{d})e^{-i\vb{k}\cdot\vb{s}}
\end{equation}
which, being a scalar, we can expand as
\begin{align}
  P(\vb{k},\vb{d}) &= \sum_{\ell} P_\ell(k,d)
  \mathcal{L}_\ell \left(\hat{k}\cdot\hat{d}\right)
\end{align}
This makes sense as, intuitively, modes comparable to the inverse distance to the galaxies correspond to widely separated pairs.
In terms of the $\xi_\ell(s,d)$ defined in \eq{eq:xiell},
\begin{align}
\label{eq:Pkell}
  P_\ell(k,d) &= 4\pi(-i)^\ell\int\,s^2\,\mathrm{d}s
  \ j_\ell(ks)\,\xi_\ell(s,d) 
\end{align}
and since the $\xi_\ell(s,d)$ have a well defined \textit{pp} limit (as we have derived in the previous section) so do the multipoles of the local power spectrum. 
As in the previous section we can work out the leading order wide angle correction to $P_\ell(k,d)$, plugging \eqs{eq:xiWA}{eq:xialpha} into the above expression. As expected, new contributions starts at order $(kd)^{-2}$ and a numerical comparison with the PP multipoles is presented in \fig{fig:PkL}.
At very large scales wide angle corrections become significant compared to the plane-parallel terms, and neglecting them could result in a non-trivial error in the modeling of power spectrum multipoles.  However by $k\simeq 0.1\,h\,{\rm Mpc}^{-1}$ the corrections are small for $\ell\le 4$. As for the correlation function, in the case of uniform samples, there is a large cancellation between the physical and the geometrical factors for $\ell=0$, $2$ that would go away with more realistic galaxy distributions.
In Fourier space all new terms will be out of phase with the Kaiser multipoles, as one can see from \eq{eq:Pkell}, but our calculations suggest their effect on the BAO would be small.
For a similar calculation in the context of general relativistic effects see the recent work of \cite{Tansella2017}.

\section{Fourier space estimators}
\label{sec:fourier_estimators}

In the above we have shown that wide angle effects are generally quite small for existing and near future experiments, though they can become a reasonable fraction of the statistical error in some cases, \eg primordial non gaussianities \citep{Dalal08,Slosar08}.
However, in these investigations we have not considered the manner in which the 2-point function is estimated.  For the configuration-space quantities, direct pair counts naturally provide the angle bisector and there is no difficulty.  For the power spectrum we must consider further how the multipoles are defined and calculated.

\subsection{Approximations in the Yamamoto estimator}

It has become standard to define the $L^{\rm th}$ multipoles of the power spectrum by integrating over pairs of points with the line of sight to each pair defined as the angle bisector or midpoint.  This leads to the Yamamoto estimator \citep{Yam06}:
\begin{align}
\label{eq:PkY}
  \hat{P}_L^Y(k) \equiv & (2L+1)\int \frac{\mathrm{d}\Omega_\vb{k}}{4\pi} \notag \\
  & \times \int \mathrm{d}^3 s_1 \mathrm{d}^3 s_2\ \delta(\vb{s}_1) \delta(\vb{s}_2) e^{-i\vb{k}\cdot \vb{s}} \mathcal{L}_L\left(\hat{k}\cdot\hat{d}\right)
\end{align}
Taking the expectation value of this estimator yields an integral over the redshift space power spectrum
\begin{align}
\avg{\hat{P}^Y_L(k)} = & (2 L+1)\int\mathrm{d}^3d\,\int \frac{\mathrm{d}^3 q}{(2\pi)^3} e^{i\vb{q}\cdot\vb{d}} \notag \\
&\times\int \frac{\mathrm{d}\Omega_{\vb{k}}}{4\pi} P(\vb{k}+\vb{q}/2,-\vb{k}+\vb{q}/2)\mathcal{L}_L(\hat{k}\cdot \hat{d})
\end{align}
which can be easily derived by working in the center-of-mass and relative separation frame \citep{Sco15} but holds true for our preferred bisector definition as well.
From the expression above it is far from clear what the relation between the Yamamoto estimator and the analytical model described in the previous section is, or in general to any theory of galaxy clustering written in the flat-sky approximation, especially in light of the difficult interpretation of the parallel limit of $P(\vb{k}_1,\vb{k}_2)$, see the discussion at the end of Sec. \ref{sec:intro}.
If we integrate over the direction of the line-of-sight and the polar angle around $\hat{d}$ we can change variables from $\vb{s}_1$ and $\vb{s}_2$ to $s$, $\mu$ and $d$ with
\begin{equation}
  \int d^3s_1\,d^3s_2 \to 4\pi\int d^2\,\mathrm{d}d
  \ 2\pi\int s^2\,\mathrm{d}s\,d\mu
\end{equation}
yielding a more familiar expression\footnote{This is true for both midpoint and angle bisector definition of the LOS.},
\begin{align}
\avg{\hat{P}_L^Y(k)}
  &=(2L+1) \int \frac{\mathrm{d}\Omega_\vb{k}}{4\pi}\, \mathrm{d}^3 d \, P(\vb{k},\vb{d}) \mathcal{L}_L(\hat{k}\cdot\hat{d}) \\
  &= \int \mathrm{d}^3 d \,P_L(k,d)
\label{eqn:PY_Pkd}
\end{align}
The Yamamoto estimator is therefore the average over all possible LOS's of the local estimate of the power spectrum, and contains wide angle corrections to $P_L(k,d)$ to all orders. 
Although we did not write it down, the above expression is intended to be normalized by the integral over the volume of the survey $\int \mathrm{d}^3 d$. In the case of simple spherical geometries the wide angle corrections will therefore controlled by $k R$ where $R$ is the size (depth) of the survey.

The estimator of \citet{Yam06} can be computationally expensive to evaluate, and for this reason an approximation is often used instead, even in the original \citet{Yam06} paper.
One replaces the line-of-sight direction, $\hat{d}$, with the direction of one member of the pair (e.g.~$\hat{s}_1$) which allows factorization of the integrals.   This is the most common assumption in the Fourier space analysis of galaxy clustering data \citep[e.g.][]{Beu14,Beu17,Ata17}. \citet{Bia15,Sco15,Han17} then showed that with this approximation the estimator could be efficiently evaluated using FFTs.

Trading $\hat{d}$ with $\hat{s}_1$ could look similar to taking the plane-parallel limit of the estimator, and therefore one would expect to make an error of the same order of the one we discussed above in Section \ref{sec:fourier_space}. However, as we will show, the bias introduced by the FFT estimator is larger that the one shown in \fig{fig:PkL}.  We start from 
\begin{align}
\label{eq:FFT}
P_L^{FFT}(k) &= (2L+1) \int \frac{\mathrm{d}\Omega_\mathbf{k}}{4\pi}\, \mathrm{d}^3 d \, P(\mathbf{k},\mathbf{s}) \mathcal{L}_L(\hat{k}\cdot\hat{s_1}) \\
&=(2L+1) \int\frac{\mathrm{d}\Omega_\mathbf{k}}{4\pi}\, \mathrm{d}^3 d \,\mathrm{d}^3 s\, \xi(s,d,\mu)\, e^{-i\mathbf{k}\cdot \mathbf{s}} \mathcal{L}_L(\hat{k}\cdot\hat{s_1}) \notag \\
&=(2L+1) \int\frac{\mathrm{d}\Omega_\mathbf{k}}{4\pi}\, \mathrm{d}^3 d \,\mathrm{d}^3 s \,e^{-i\mathbf{k}\cdot \mathbf{s}}\notag \\
& \times\sum_{\ell} \xi_\ell(s,d)\mathcal{L}_\ell(\hat{s}\cdot\hat{d})\mathcal{L}_L(\hat{k}\cdot\hat{s}_1) \\
&=(2L+1) (4\pi)(-i)^L\int\mathrm{d}^3 d \,\mathrm{d}^3s
\ j_L(ks) \notag \\
& \times\sum_{\ell} \xi_\ell(s,d)\mathcal{L}_\ell(\hat{s}\cdot\hat{d})\mathcal{L}_L(\hat{s}\cdot\hat{s}_1)
\end{align}
Using $\hat{s}\cdot\hat{s}_1=\mu+(1/2)(1-\mu^2)x+\cdots$ to expand $\mathcal{L}_L(\hat{s}\cdot\hat{s}_1)$ in $x$ and $\mu$, similarly to what we did in Section \ref{sec:linear_theory}, we find, at leading order,
\begin{align}
\label{eq:PkFFT}
\avg{\hat{P}_L^{FFT}(k)}= \int \mathrm{d}^3 d\, \left[ P_L^{(2)}(k,d) +  \left(\frac{1}{kd}\right)^2\sum_i b_{iL}\mathcal{P}^{L}_i(k) \right]
\end{align}
with $P_L^{(2)}(k,d)$ the power spectrum multipoles computed up to $\mathcal{O}[(kd)^{-2}]$, as for instance in the previous section within linear theory, the $b_{iL}$ are constant coefficients defined in \eq{eq:bij} and
\begin{equation}
\mathcal{P}^{L}_i(k) \equiv  4 \pi(-i)^L \int \mathrm{d}s\,s^2\,(k s)^2 j_L(ks)\xi_i(s,x=0)
\end{equation}
where by $\xi_k(s,x=0)$ we mean the configuration space multipoles in the plane parallel limit. In linear theory they would correspond to \eq{eq:xiK}, but the above result is valid at any order in perturbation theory of the density field.

\begin{figure}
\centering
\resizebox{\columnwidth}{!}{\includegraphics{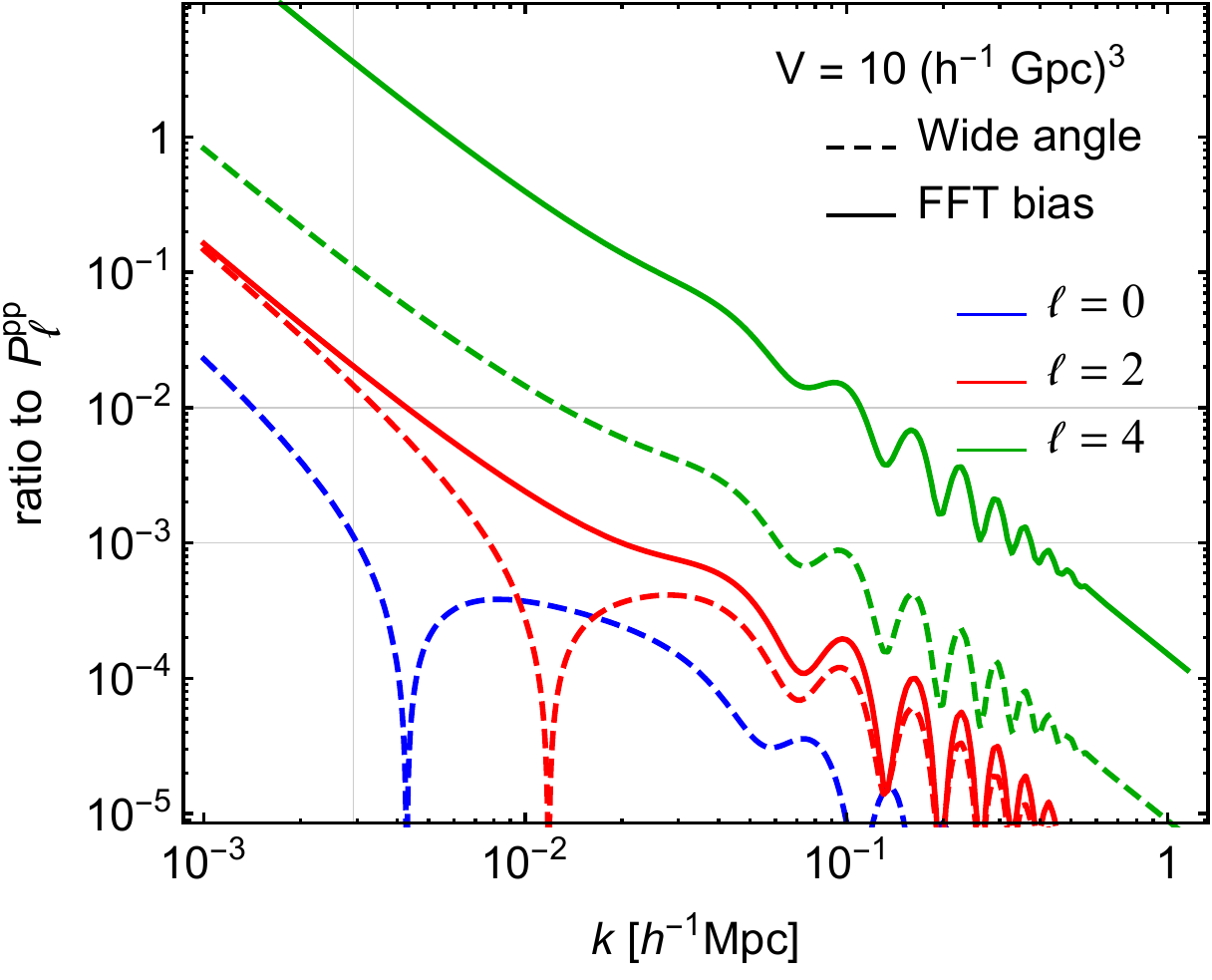}}
\caption{Error introduced by an incorrect definition of the LOS in the estimator in \eq{eq:FFT}. For comparison the intrinsic wide angle terms of \fig{fig:PkL} are also shown with dashed lines.}
\label{fig:FFT}
\end{figure}

To the best of our knowledge this is the first time corrections to the FFT estimator have been computed analytically in the most general case. Previous work in \cite{Samushia15} discussed the limits of the FFT estimator, but in a simplified set up.
What we find is that the assumption $\hat{s}\simeq\hat{s}_1$ mixes Hankel transforms on different multipoles in a very non-trivial way.
For $L=0$ there are no extra corrections, recovering the  well known result that for the monopole the Yamamoto estimator is identical the one in \citet{FKP}. For $L\ne0$ the remaining coefficients are
\begin{align}
\label{eq:bij}
b_{02} &= \frac{3}{4}\;,\;b_{22} = -\frac{9}{28}\;,\;b_{24} = \frac{2}{21} \notag \\
b_{04} &= 0\;,\;b_{42} = \frac{15}{14}\;,\;b_{44} = -\frac{185}{154}
\end{align}
If correlation function multipoles with $\ell>4$ are present, for instance because of nonlinearities, they will also generate new terms in \eq{eq:PkFFT}. The same is true for odd $\ell$, \eg relativistic dipoles. For simplicity we will assume only $\ell=0,2,4$ are different from zero.
\fig{fig:FFT} shows the ratio between the new terms introduced by the FFT estimator and the PP multipoles. Our example describes a full sky survey with volume $V =10 \Gpc$ with mean redshift $z=1$. The vertical line in the figure represents the fundamental mode of the survey. 
The bias introduced by the FFT estimator is comparable to, or larger than, the one induced by physical wide angle terms and it results only from an incorrect choice of the line of sight. It is also strongly dependent on $\ell$, with the error on the hexadecapole much bigger than the one on the quadrupole.

\begin{figure}
\centering
\resizebox{\columnwidth}{!}{\includegraphics{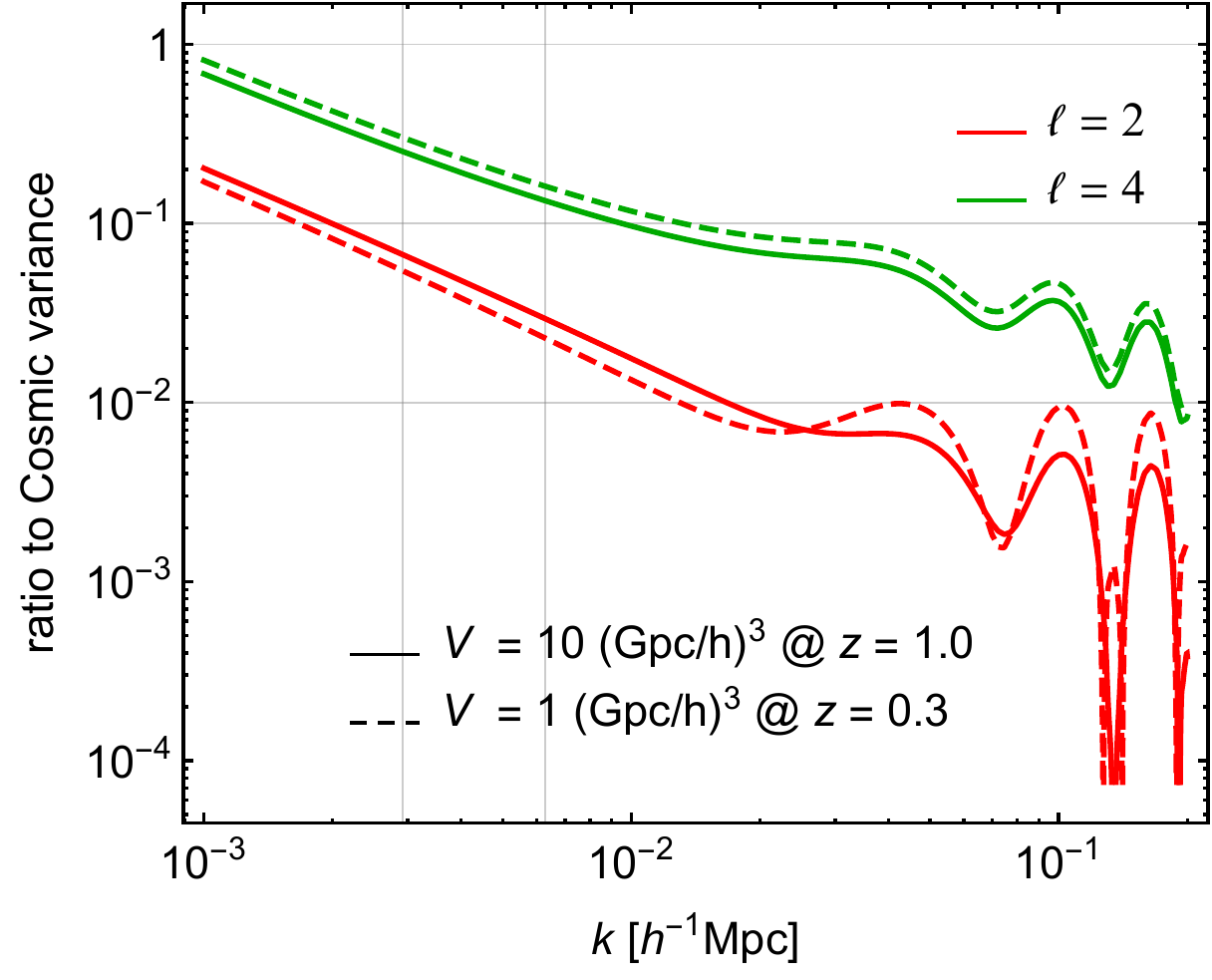}}
\caption{Comparison between the cosmic variance error in a measurement of the power spectrum multipoles and the wide angle effects, including ones introduced by the estimator in \eq{eq:FFT}. Different colors represent differenct multipoles, and dashing indicate the mean redshift of the hypotetical measurement.}
\label{fig:CV}
\end{figure}

Finally let us comment on the error budget of a measurement of the power spectrum.
In data, large scales are the ones mostly affected by cosmic variance and therefore have the largest error bars. For redshift space multipoles one has, under the Gaussian approximation,
\begin{align}
\sigma^2_{P_\ell} = \frac{2}{N_k}\frac{(2\ell+1)^2}{2}\int \mathrm{d}\mu_k \,\mathcal{L}_\ell(\mu_k)^2 P(k,\mu_k)^2
\end{align}
where we have assumed shot noise is negligible, the number of modes is defined as
\begin{equation}
N_k \equiv \frac{4 \pi k^2 \Delta k}{(2\pi)^3} V
\end{equation}
and the $k$-binning is assumed to be constant.  We will take $\Delta k= 5 \times 10^{-3} \kMpc$ as in \citet{Beu14}. In this simplified case it is easy to see that the error on the multipoles scales as 
\begin{equation}
\sigma_{P_\ell} \propto \frac{1}{[k(\Delta k R^3)^{1/2}]}\,P(k)
\qquad ,
\end{equation}
which means the ratio between the wide angle terms in \eq{eq:PkFFT} and cosmic variance scales as $(k^2/\Delta k \times R)^{-1/2}$.  This means wide angle corrections will always be smaller than the cosmic variance, although at the largest scales, when $kR\simeq 1$, the two become comparable. In \fig{fig:CV} we show the ratio between the sum of all wide angle corrections and cosmic variance. We plot two different configurations, a $V= 1 \Gpc^3$ survey at $z=0.3$ with dashed lines and a $V = 10 \Gpc^3$ one at $z=1$ with continuous lines. These numbers are similar to what a typical redshift bin of a BCG sample and ELG sample in DESI could look like \citep{DESI}. For the quadrupole the error is always negligible compared to cosmic variance, but for the hecadecapole can contribute up to 20-30\% on the largest scales.
If one is able to (partially) cancel sample variance, using for instance cross-correlations \citep{Seljak09,MdDonaldSeljak09}, then wide angle effects may become the main source of theoretical systematics.

The formalism developed above suggests one strategy for efficiently dealing with these effects.  The errors introduced are only significant at relatively large scales, where linear theory is a good approximation.  Within linear theory calculating the corrections to the theoretical predictions is straightforward for any of the statistics.  Thus forward modeling the correction terms along with the theoretical prediction should be an easy and efficient strategy.

\subsection{Wedges}

While the multipole expansion is the most common choice of statistic, some authors instead use ``wedges'' (i.e.~bins) in $\mu_k$ \citep{Grieb17,Sanchez17}.  One advantage of wedges over multipoles is that systematics may be localized in a particular $\mu_k$ wedge, and in such situations one can just throw that bin away without affecting the rest of the data.  Examples of systematic effects with such an angular structure are foreground contamination \citep{Colavincenzo} and fiber assignment in spectroscopic surveys \citep{Hahn17,Burden17,Pinol17}, which both live at $\mu_k\simeq 0$. For instance, \citet{Han17} show how non-uniform wedges can mitigate the effect of fiber assignment for the DESI experiment.

In the flat sky limit estimating $P(k,\mu_k)$, and hence the wedges, is straightforward.  But in the more general case there is no unambiguous way to define the parallel, $k_{\parallel}$, and perpendicular, $k_\perp$, components of the 3D vector $\vb{k}$ for the entire survey. One solution is to estimate several $P_L(k)$ and use the fact that Legendre polynomials form a complete basis in [-1,1]
\begin{equation}
  P(k,\mu_k) = \sum_{L=0}^{L_{\rm max}} P_L(k) \mathcal{L}_L(\mu_k)
\end{equation}
In linear theory $L_{\rm max} = 4$, however non linearities and systematic effects can contribute to higher multipoles.  The disadvantage is that constructing narrow bins in $\mu_k$ requires a very large value of $L_{\rm max}$.  \citet{Han17} advocate $L_{\rm max}=16$ for example.  As estimating many multipoles can require a lot of time and memory the use of FFT estimators, like the one in \eq{eq:PkFFT}, is crucial. 
Our analysis suggests that such FFT estimators will become more and more biased with respect to the full Yamamoto case (Eq.~\ref{eq:PkY}) as $L$ increases. Although the effect will depend on the structure of the particular systematic effect one is interested in, we caution against the use of the estimator in \eq{eq:PkFFT} without a careful comparison with the correct answer in \eq{eq:PkY}. Our formulae provide the analytic framework to study these effects, and we plan to return to some specific examples in future work.

In Section \ref{sec:sFB} we will discuss new estimators that naturally isolate low-$k$ modes without the penalty of measuring high-$L$ multipoles.

\subsection{The effect of masks}
\label{sec:Pkmask}
We now consider the impact of the survey geometry on our estimates of $P(k)$.  In the standard approach the survey mask acts multiplicatively on the density field in configuration space, and thus the measured Fourier modes are convolutions of the density and window Fourier transforms.  For diagonal power spectra (Eq.~\ref{eqn:plane_parallel_P}) one obtains the standard result that
\begin{equation}
  P(\vb{k}) = \int \frac{d^3q}{(2\pi)^3}
  \ P(\vb{q})\left| W(\vb{k}-\vb{q}) \right|^2
  \quad .
\label{eqn:standard_window}
\end{equation}
The impact on the multipoles is then found by expanding the window functions in multipole moments.
However in deriving Eq.~(\ref{eqn:standard_window}) we have made assumptions which are violated in the case of wide angles where we have lost the translational invariance, which is key to the above. This equation is indeed valid only if a global line of sight can be defined for each pair of galaxies, $\hat{d}\rightarrow \hat{z}$, and we know this is a bad approximation for surveys covering a large fraction of the sky. In fact the situation is more similar to the impact of masks on pseudo-$C_\ell$ estimators of angular power spectra, used frequently in CMB research \citep[e.g.][]{Hiv02,P13_XV,Els17}.
In this section we therefore derive the effect of the mask on the Yamamoto estimator. Let's start with 
\begin{align}
\avg{\hat{P}_L(k)} = &(2L+1)\int \frac{\mathrm{d}\Omega_\vb{k}}{4\pi} \int \mathrm{d}^3 s_1 \mathrm{d}^3 s_2 \notag \\
\times & W(\vb{s}_1)W(\vb{s}_2)\,\xi(\vb{s}_1,\vb{s}_2) e^{-i\vb{k}\cdot \vb{s}} \mathcal{L}_L\left(\hat{k}\cdot\hat{d}\right) \notag \\
 = & (2L+1)\int \frac{\mathrm{d}\Omega_\vb{k}}{4\pi} \int \mathrm{d}^3 s \,\mathrm{d}^3d\ e^{-i\vb{k}\cdot \vb{s}} \notag \\
 \times &  W^+(\vb{s},\vb{d})W^-(\vb{s},\vb{d}) \xi(\vb{s},\vb{d}) \mathcal{L}_L\left(\hat{k}\cdot\hat{d}\right)
\end{align}
where we have defined 
\begin{align}
W^{+}(\vb{s},\vb{d})\equiv W[\vb{s}_1(\vb{s},\vb{d})]\;,\; W^{-}(\vb{s},\vb{d})\equiv W[\vb{s}_2(\vb{s},\vb{d})] 
\end{align}
Next we Fourier transform the window functions with respect to $\vb{s}$ and perform the angular integral, $d\Omega_k$,
\begin{align}
\avg{\hat{P}_L(k)} =& (2L+1) \int \mathrm{d}^3 s \,\mathrm{d}^3d  \int \frac{\mathrm{d}^3 k_1}{(2 \pi)^3} \,\frac{\mathrm{d}^3 k_2}{(2 \pi)^3}  \,e^{i\vb{k}_1\cdot \vb{s}}   e^{i\vb{k}_2\cdot \vb{s}}\notag \\
& (-i)^L \sum_\ell \xi_\ell(s,x)\mathcal{L}_\ell(\hat{s}\cdot\hat{d})\mathcal{L}_L(\hat{s}\cdot\hat{d}) \,j_L(ks) \notag \\
& W^{+}(\vb{k}_1,\vb{d}) W^{-}(\vb{k}_2,\vb{d})
\label{eq:simpler_window_expression}
\end{align}
which we rewrite as
\begin{align}
\avg{\hat{P}_L(k)} = & (2L+1)\sum_{\ell}
\sum_{\substack{L_1,L_2,L_3\\M_1,M_2,M_3}}
\int s^2\mathrm{d} s\,\mathrm{d}^3d\ Y^\star_{L_1,M_1}(\hat{d}) \notag \\
& \times (4\pi)^3\tj{l}{L}{L_1}{0}{0}{0}^2 \mathcal{G}^{M_1,M_2,M_2}_{L_1,L_2,L_3} \,\xi_\ell(s,x) j_L(ks)\notag \\
& \times \int \frac{\mathrm{d}^3 k_1}{(2 \pi)^3} \,\frac{\mathrm{d}^3 k_2}{(2 \pi)^3}  W^{+}(\vb{k}_1,\vb{d}) W^{-}(\vb{k}_2,\vb{d}) i^{L_2+L_3-L}
\notag \\& \times j_{L_2}(k_1 s) j_{L_3}(k_2 s)Y^\star_{L_2,M_2}(\hat{k}_1)Y^\star_{L_3,M_3}(\hat{k}_2)
\end{align}
where
\begin{align}
  \mathcal{G}^{M_1,M_2,M_2}_{L_1,L_2,L_3} &=
  \sqrt{\frac{(2L_1+1)(2L_2+1)(2L_3+1)}{4\pi}}  \notag \\
  &  \tj{L_1}{L_2}{L_3}{0}{0}{0}\tj{L_1}{L_2}{L_3}{M_1}{M_2}{M_3}
\end{align}
is the Gaunt integral.  We notice that the last two lines in the equation are simply the Fourier-Bessel transforms of the window function,
\begin{align}
\int\frac{\mathrm{d}^3k}{(2\pi)^3} Y_{\ell m}^\star(\hat{k}) j_\ell(ks)\,W(\vb{k},\vb{d}) &=
\frac{(-i)^\ell}{4\pi}\int \mathrm{d}\Omega_{s} Y_{\ell m}^\star(\hat{s})W(\vb{s},\vb{d}) \notag \\
&= \frac{(-i)^\ell}{4\pi}\ W_{\ell m}(s,\vb{d})
\end{align}
so that
\begin{align}
\avg{\hat{P}_L(k)} =  & (2L+1)\sum_{\ell} \sum_{\substack{L_1,L_2,L_3\\M_1,M_2,M_3}} \int s^2\mathrm{d} s\,\mathrm{d}^3d\ Y^\star_{L_1,M_1}(\hat{d}) \notag \\
& \times\tj{l}{L}{L_1}{0}{0}{0}^2 \mathcal{G}^{M_1,M_2,M_2}_{L_1,L_2,L_3} \,\xi_\ell(s,x) j_L(ks)\notag \\&
\times (4\pi)(-i)^L W^+_{L_2,M_2}(s,\vb{d})W^-_{L_2,M_2}(s,\vb{d})
\end{align}
As expected, since the correlation function does not depend on the orientation of the galaxy pairs, the final results depend only on the length of the separation vectors, $s$, and we can sum over the $M$'s. This expression, as anticipated, looks fairly similar to the pseudo-$C_\ell$ analysis of CMB data. 
In a more compact form we can write
\begin{align}
\label{eq:PkW}
\avg{\hat{P}_L(k)} =&  4\pi (-i)^L (2L+1) \sum_{\ell}\int d^2\mathrm{d}d\,s^2\mathrm{d}s \notag \\
& \ \xi_\ell(s,x) j_L(ks)\widetilde{W}_\ell^L(s,d)
\end{align}
with
\begin{align}
 \widetilde{W}_\ell^L(s,d)\equiv &  \sum_{\substack{L_1,L_2,L_3\\M_1,M_2,M_3}} \int \mathrm{d}\Omega_{\vb{d}}\ Y^\star_{L_1,M_1}(\hat{d}) \tj{l}{L}{L_1}{0}{0}{0}^2 \notag \\ & \times  \mathcal{G}^{M_1,M_2,M_2}_{L_1,L_2,L_3} W^+_{L_2,M_2}(s,\vb{d})W^-_{L_2,M_2}(s,\vb{d})
\end{align}
Note for a full-sky survey the integral over $\Omega_d$ sets all of the $L_i=M_i=0$ and thus the window is non-zero only when $\ell=L$ (where it equals unity and we regain the expressions of \S\ref{sec:fourier_estimators}). However in general the mask will mix different Hankel transforms of the correlation function multipoles, in contrast with the standard assumption of \eq{eqn:standard_window}, which yields an expression with $\ell=L$ \citep{Beu14,Han17}.
The importance of the new terms in \eq{eq:PkW} will strongly depend on the detailed shape of the window function of the specific galaxy survey, and so we defer such an analysis to future work.

We note that at the practical level it is slightly easier to define $\widetilde{W}_\ell^L$ via
\begin{align}
\label{eq:WlL}
\widetilde{W}_\ell^L(s,d) = & \frac{1}{4\pi}\int d\Omega_d\, d\Omega_s  \ \mathcal{L}_\ell(\hat{s}\cdot\hat{d})\mathcal{L}_L(\hat{s}\cdot\hat{d}) \notag \\
& W^{+}(\vb{s},\vb{d}) W^{-}(\vb{s},\vb{d})
\end{align}
which follows from \eq{eq:simpler_window_expression} upon undoing the Fourier transforms and comparing to \eq{eq:PkW}.  The window function so-defined can be constructed by taking a random catalog and, for each pair of randoms, constructing the angle bisector and separation vector.  Summing over pairs and binning the integrand in $s$ and $d$ gives a Monte-Carlo evaluation of the angle integrals defining $\widetilde{W}_\ell^L(s,d)$.  \eq{eq:PkW} may then be put in a more familiar form by writing $\xi_\ell(s,x)$ as the Hankel transform of $P_\ell(k,d)$ and grouping the remaining terms into a window function depending on $L$, $\ell$, $k$ and $k'$.
The results presented in this Section concern the full Yamamoto estimator in \eq{eq:PkY}, for which the calculation of the masked power spectrum is actually easier than for the FFT estimator in \eq{eq:FFT}. The main reason is that the Legendre polynomials in \eq{eq:WlL} have the same argument only in the Yamamoto estimator, whereas in the FFT case, the one most used in data analysis, one would have to deal with extra terms in $\mu$ and $x$. From a comparison of \eq{eq:FFT} and \eq{eq:simpler_window_expression} we believe it should be clear how to proceed in the latter case.  
A similar computation of the effect of the mask on 3D multipoles of the power spectrum appeared in Appendix A of \cite{Beu17}. There are two main differences between the latter and this work. First of all we compute the effect of the mask for the exact Yamamoto estimator, as the difference with respect to the FFT one will just introduce new multipole moments of the correlation function, see \eq{eq:PkFFT}. Secondly and most importantly we compute the convolution of the theoretical models with window functions including leading order wide angle corrections, which are important roughly at the same scale the mask is and therefore cannot be neglected. 
We also note that \eqs{eq:PkW}{eq:WlL} are the correct expressions to use if the redshift evolution of the clustering signal has to be taken into account, \eg if the power spectrum is measured in wide redshift bins.

\section{Spherical Fourier Bessel Analysis}
\label{sec:sFB}

Many of the issues encountered above arise because the symmetries of the Fourier transform are not well matched to the symmetries of the survey.  It is thus natural to use a different transformation, one which naturally respects the split into radial and angular modes.  Such a transformation is the well known spherical Fourier-Bessel (sFB) expansion. Since the original paper of \cite{HT95} several authors have studied galaxy clustering in spherical coordinates, see for instance \cite{YD2013,Pratten13,Nic14,Liu16,Pas17} and references therein.
These analyses retain a clear separation between angular and radial coordinates, \ie redshifts. The methods have been successfully applied to data in \citet{Fisher94,Tadros99,Tay01,Pad01,Percival04,Pad07}. For a bipolar harmonic analysis see also the recent work by \cite{Shiraishi17}.

\subsection{Formalism}

If we define the Fourier transform of a density field in configuration space, $\delta(\vb{r})$, as
\begin{equation}
\delta(\vb{k}) = \int \mathrm{d}^3s\, \delta(\vb{s}) e^{-i \vb{k}\cdot\vb{s}} 
\end{equation}
then we can define the forward and backward sFB transforms as 
\begin{equation}
\label{eq:sFB}
\delta_{\ell m}(k) = \sqrt{\frac{2}{\pi}}\int \mathrm{d}^3s \ Y_{\ell m}^\star(\hat{s}) j_\ell(ks) \,\delta(\vb{s})
\end{equation}
\begin{equation}
\delta(\vb{s}) = \sum_{\ell m} \sqrt{\frac{2}{\pi}}\int k^2\mathrm{d}k\ Y_{\ell m}(\hat{s}) j_\ell(ks) \,\delta_{\ell m}(k) 
\end{equation}
where $Y_{\ell m}(\hat{s})$ are spherical harmonics and $j_\ell(ks)$ spherical Bessel functions.
It is important to notice that a-priori there is no reason for the wave vector in \eq{eq:sFB} to have the same magnitude of  the 3D vector $\vb{k}$ associated with the Fourier transform of the density field. While the former is associated with radial modes, the latter lives in Cartesian coordinates.
A convenient intermediate step is to define angular coefficients similar to the CMB case
\begin{equation}
\label{eq:alm}
\delta(\vb{s}) =  \sum_{\ell m} a_{\ell m}(s) Y^{\star}_{\ell m}(\hat{s})
\end{equation}
and then Hankel-transform the $a_{\ell m}$'s with respect to $s$. 
The two point function of the density field is simply related to that of the configuration space multipoles 
\begin{align}
 \xi^s(\vb{s}_1 , \vb{s}_2) &= \sum_{\ell m} \avg{a_{\ell m}(s_1)a^\star_{\ell m}(s_2)} Y_{\ell m}(\hat{s}_1)Y^\star_{\ell m}(\hat{s}_2)\notag \\
 &\equiv \sum_{\ell m} C_\ell(s_1, s_2) Y_{\ell m}(\hat{s}_1)Y^\star_{\ell m}(\hat{s}_2) \notag \\
 &=\sum_\ell\frac{2\ell+1}{4\pi}C_\ell(s_1,s_2)\mathcal{L}_\ell(\hat{s}_1\cdot\hat{s}_2)
\label{eq:xi_from_Cl}
\end{align}
A straightforward calculation shows that in linear theory
\begin{align}
\label{eq:Csell}
 C_\ell(s_1, s_2) =& \frac{2}{\pi} \int \mathrm{d}\,k\,k^2 P(k) \notag \\
 &[j_\ell(ks_1)-f(z)j_\ell''(ks_1)][j_\ell(ks_2)-f(z)j_\ell''(ks_2)]
\end{align}
where $j''(x)$ is the second derivative of the $j$'s with respect to its argument. We can relate the $C_\ell(s_1,s_2)$ to the 3D multipoles $\xi_L(s,x)$, from now on indicated with a capital letter $L$, with the help of \eq{eq:xiell}.
In the small angle limit,
\begin{align}
\xi_L(s,x) \simeq \sum_\ell \int \mathrm{d} \mu\,&  \frac{2\ell+1}{2 \pi(2L+1)}  C_\ell(s_1,s_2) \notag \\
&\times \mathcal{L}_\ell\left(1-\frac{[1-\mu^2] x^2}{4}\right) \mathcal{L}_L(\mu)
\end{align}
and a similar inverse relation exists\footnote{To obtain the inverse relation, integrate \eq{eq:xi_from_Cl} over $d(\cos\theta)$ times a Legendre polynomial in $\cos\theta$ and express $\xi^{s}(\vb{s}_1,\vb{s}_2)$ in terms of $x$, $d$ and $\mu$.}.
Note that since $\ell$ can be arbitrarily large we are not allowed to expand the Legendre polynomial in $\ell$, as we did before, since $\ell x$ can easily be $\mathcal{O}(1)$.  However we can use $\mathcal{L}_\ell(\hat{s}_1\cdot\hat{s}_2)\simeq J_0(\ell\,x\sqrt{1-\mu^2})$ (see Appendix \ref{app:identities}).  Under this approximation we recognize the sum over $\ell$ as the Fourier transform for an azimuthally symmetric function in angular coordinates, and the argument of $J_0$ as $k_\perp s_\perp$ with $\ell=k_\perp d$.  If we were to further Fourier transform on $s_\parallel$ we would obtain the standard expression for $\xi(\vb{s})$ as the Fourier transform of $P(k_\parallel,k_\perp)$.  Note that the $C_\ell(s_1,s_2)$ are simply related to the ``MAPS'' of \citet{Dat07}, as there exists a one-to-one mapping between frequencies of an emitted signal and redshift/distances.

Let's now move to the Fourier analysis.
In real space we know the density field is statistically homogeneous and isotropic, which means the power spectrum of sFB coefficients does not depend on $\ell$ nor on $m$ but just on the magnitude of the wave-vector $k$
\begin{equation}
\label{eq:sFBr}
\avg{\delta_{\ell_1m_1}(k_1) \delta^\star_{\ell_2m_2}(k_2)} = P(k)
\frac{\delta_D(k_1-k_2)}{k^2}\delta^K_{\ell_1 \ell_2} \delta^K_{m_1 m_2}
\end{equation}
where $\delta^K_{i,j}$ is a Kroenecker delta. As a result of the symmetry of the problem, the 1D $k$ of the sFB transform coincides with the 3D one in $P(k)$.
On the other hand in redshift space we broke translational and partially rotational invariance, so we expect the sFB power spectrum to be independent only of $m$, the eigenvalue associated with the remaining azimuthal symmetry,
\begin{equation}
\label{eq:sFBRSD}
\langle \delta_{\ell_1 m_1}(k_1) \delta^\star_{\ell_2 m_2}(k_2) \rangle = C_\ell(k_1,k_2) \delta^K_{\ell_1 \ell_2} \delta^K_{m_1 m_2} 
\end{equation}
In this case there is no simple relation between the 3D modes and the radial modes on the RHS of the above equation.
By Hankel-transforming \eq{eq:Csell} twice it is easy to see that in linear theory the redshift space angular multipoles are
\begin{align}
C_\ell(k_1,k_2) = & \int \mathrm{d} k \,k^2 P(k)  \int \mathrm{d}s_1\,s_1^2\,\int \mathrm{d}s_2\,s_2^2
\,j_{\ell}(k_1 s_1)j_{\ell}(k_2 s_2) \notag \\
& [j_{\ell}(ks_1)-f j_{\ell}''(ks_1)]
  [j_{\ell}(ks_2)-f j_{\ell}''(ks_2)]
\end{align}
We simplify this expression in Appendix \ref{app:lin_expression}.

Beyond the linear regime, most, if not all, analytic models of structure formation are built within the flat sky approximation, as in this case there is a well defined angle between the Fourier mode and the LOS. On the other hand, as we have seen, estimating power spectrum multipoles forces the inherently non-flat geometry of the curved sky to become a 3D Cartesian system.
This brings in a few shortcomings, depending on how many approximations one is willing to take, as we have discussed in Sec. \ref{sec:fourier_estimators}.
The ideal basis to estimate redshift space clustering would therefore be a spherical one, \eg of sFB coefficients, which does not require defining a 3D wave vector $\vb{k}$.

However, the major obstacle to the use of sFB power spectra is its relation to theoretical models. As we have seen in \eq{eq:sFBRSD} there is no straightforward way of relating the angular multipole $\ell$ to $\mu_k$, and the loss of translational invariance implies the $C_\ell(k_1,k_2)$ depends on both radial wavevectors. This tremendously  complicates perturbation theory approaches beyond the linear regime and estimating the covariance matrix becomes incredibly more difficult.
The bottom line therefore seems to be that the basis in which theory is more well understood, \ie Cartesian 3D coordinates, and the one more suited for the measurement, \eg sFB's, are not the same.

The degree to which this is a practical problem remains to be seen.  In general non-linear corrections are important only on small scales, where wide-angle effects are typically small.  One could imagine a hybrid strategy in which linear theory is used to compute $C_\ell(k_1,k_2)$ for small $k_i$ with a smooth switch to non-linear models in which the flat-sky limit has been assumed to translate from $P(k_\perp,k_\parallel)$.
However this will work only at the level of the theoretical modeling and not for the estimator, which is the first place in a data analysis where we want to have good control of wide angle effects. 
In the next Section we will show how to solve this problem using sFB coefficients.

\subsection{FFT estimator in spherical coordinates}
\label{sec:PksFB}

The discussion above motivates the study of estimators for the 3D power spectrum written in terms of sFB coefficients. As we have seen in \S\ref{sec:fourier_estimators}, if one restricts the analysis to low multipoles (up to the hexadecapole) the error introduced by the FFT estimator in the comparison with a PP theory is small.
However observational systematics sometimes require the computation of the multipoles to a much higher order to be properly removed, or a well motivated template to be marginalized over \citep{Hahn17}.
As we will show in the next few lines sFB coefficients have the very appealing property of naturally accounting for systematics at $\mu=0$ even starting from the FFT estimator in \eq{eq:PkFFT}. We first rewrite the FFT estimator, \eq{eq:PkY}, as
\begin{align}
&\hat{P}_L^{FFT}(k) = (4\pi)^3\sum_{\substack{\ell_1m_1\\ \ell_2 m_2}}\sum_M i^{\ell_1-\ell_2}\int \frac{\mathrm{d}\Omega_\mathbf{k}}{4\pi} Y_{LM}(\hat{k}) \times
\notag \\ & 
\int \mathrm{d}^3 s_1 \mathrm{d}^3 s_2 \ \delta(\vb{s}_1) \delta(\vb{s}_2)j_{\ell_1}(k s_1)j_{\ell_2}(k s_2) \times \notag \\& Y_{\ell_1m_1}(\hat{k})Y^\star_{\ell_1m_1}(\hat{s_1})Y^\star_{LM}(\hat{s_1}) Y_{\ell_2m_2}(\hat{k})Y^\star_{\ell_2m_2}(\hat{s_2}) 
\end{align}
and then integrate over $d\Omega_k$ to arrive at
\begin{align}
&\hat{P}_L^{FFT}(k) =(4 \pi)^3 \sum_{\ell m} \sum_{\ell_1m_1}\sum_{\ell_2m_2}\sum_M i^{\ell_1-\ell_2} \mathcal{G}_{\ell_1 \ell_2 L}^{m_1m_2M} \mathcal{G}_{\ell_1 \ell L}^{m_1m M} \notag \\
&\times \int \mathrm{d}^3 s_1 \mathrm{d}^3 s_2 \ \delta(\vb{s}_1) \delta(\vb{s}_2) Y_{\ell m}(\hat{s_1}) Y^\star_{\ell_2m_2}(\hat{s_2}) j_{\ell_1}(k s_1)j_{\ell_2}(k s_2)
\end{align}
where we used the definition of the Gaunt integral twice. With the help of \eq{eq:threejI} we can simplify the above expression significantly to obtain
\begin{align}
&\hat{P}_L^{FFT}(k) =  (4 \pi)(2L+1)  \notag \\
&\sum_{\ell m}  \sum_{\ell_1<|\ell-L|}^{\ell+L} i^{\ell_1-\ell} (2\ell _1+1) \tj{\ell}{\ell_1}{L}{0}{0}{0}^2 \times\notag \\
&\int \mathrm{d}^3s_1\, \mathrm{d}^3 s_2\, \delta(\vb{s}_1) \delta(\vb{s}_2) Y_{\ell m}(\hat{s}_1) Y^\star_{\ell m}(\hat{s}_2) j_{\ell_1}(k s_1) j_{\ell}(k s_2) \notag \\
=& (4 \pi)(2L+1)\sum_{\ell m}  \sum_{\ell_1<|\ell-L|}^{\ell+L} i^{\ell_1-\ell} (2\ell_1+1) \tj{\ell}{\ell_1}{L}{0}{0}{0}^2 \times \notag \\
&\int \mathrm{d}s_1\,s_1^2\, \mathrm{d}s_2\,s_2^2\,a_{\ell m}(s_1) a_{\ell m}^\star(s_2) j_{\ell_1}(k s_1) j_{\ell}(k s_2)
\end{align}
We recognize in the above expression the spherical harmonic expansion of the density field, but they are coupled to spherical Bessel function of a different order. We therefore define the generalized sFB coefficients 
\begin{equation}
\delta_{\ell m}^L(k) = \sqrt{\frac{2}{\pi}}\int \mathrm{d}^3s\,  \delta(\mathbf{s})\,Y_{\ell m}(\hat{s}) j_{L}(k s)
\end{equation}
which form an overcomplete basis. The final expression for the FFT estimator is
\begin{align}
\hat{P}_L^{FFT}(k) =& (2 \pi^2)(2L+1)\sum_{\ell m}  \sum_{\ell_1<|\ell-L|}^{\ell+L}i^{\ell_1-\ell} \notag \\
& (2\ell_1+1) \tj{\ell}{\ell_1}{L}{0}{0}{0}^2 \delta_{\ell m}^{\ell_1}(k)\delta_{\ell m}^\star(k)
\label{eq:PY_sFB}
\end{align}
Since $L$ is even, $\ell+\ell_1$ is also even, and the power spectrum is therefore positive. On the other hand the above equation also provides a clear way to estimate imaginary parts like dipoles, \eg \cite{Gaztanaga17}.
The important feature of the above estimator, similar to \eq{eq:sFBr}, is that the length of the 3D vector $\vb{k}$ on the right hand side and the 1D radial wavenumber appearing as the argument in the sFB coefficients are the same. 
This means that if one wants to discard the low $k_\parallel$ modes because of contamination by systematics, it is enough to omit the first few sFB coefficients.  This will automatically remove them from the 3D multipoles, $P_L(k)$. In contrast to the method in \citet{Han17}, our approach works at the level of the field and it does not rely on the detailed structure of the systematic at hand. 
A similar expression for the bispectrum is presented in the Appendix. For higher point functions the technique in \cite{Han17} does not apply anymore, whereas \eq{eq:PY_sFB} can still be used to remove systematics in the plane of the sky.
In terms of computational cost it should be kept in mind that spherical transforms are more expensive than FFTs, scaling as $N^{3/2}$ for the angular part, where $N$ is the number of pixels in the map, and as $l_{\rm max} \times N_k \log N_k$ in the radial direction. However one could imagine using the symmetry of the problem, \eg $m$-independence, to speed up the evaluation of sFB coefficients.

\subsection{The mask in spherical estimators}

In Section \ref{sec:Pkmask} we noticed that, contrary to what is usually assumed, the estimators for the multipoles of the power spectrum cannot be written as a simple convolution in the presence of a mask. To correctly account for incomplete sky coverage we had to define two different masks, now functions of the separation between each pair of galaxies. This is not a problem per se, but it certainly requires additional computational resources. In a spherical analysis there is no need to define multiple window functions. In the presence of an angular mask, $W(\hat{s})$, and a radial selection function, $\phi(s)$, we define the generalized sFB coefficients as
\begin{align}
\tilde{\delta}_{\ell m}^L(k) \equiv &\sqrt{\frac{2}{\pi}} \int \mathrm{d}^3s\, [\phi(s)W(\hat{s})] \delta(\mathbf{s})\,Y_{\ell m}(\hat{s}) j_{L}(k s) \notag \\
 =&\sqrt{\frac{2}{\pi}}  \sum_{\ell_1 m_1}\sum_{\ell_2 m_2}  \mathcal{G}_{\ell \ell_1 \ell_2}^{m m_1 m_2} W_{l_2m_2} \notag \\
&\left[\int \mathrm{d}s\,s^2\,a_{\ell_1 m_1}(s)j_{L}(k s) \phi(s)\right] \notag \\
 \equiv &  \sum_{\ell_1 m_1}\sum_{\ell_1 m_2}\mathcal{G}_{\ell \ell_1 \ell_2}^{m m_1 m_2} W_{l_2m_2} \Delta_{\ell_1m_1}^L(k)
\end{align}
where $W_{\ell m}$'s are the harmonic coefficients of the angular mask, and $\Delta_{\ell_1m_1}^L$ is the Hankel transform of $a_{\ell_1m_1}(s)\phi(s)$.
At the power spectrum level we obtain
\begin{align}
\sum_m\avg{\tilde{\delta}_{\ell m}^L(k) \tilde{\delta}_{\ell m}^\star(k)} &= \sum_m\sum_{\ell_1 m_1}\sum_{\ell_2 m_2}\sum_{\ell_3 m_3}\sum_{\ell_3 m_4} W_{\ell_2m_2}W_{\ell_4m_4} \notag \\
&\times  \mathcal{G}_{\ell \ell_1 \ell_2}^{m m_1 m_2} \mathcal{G}_{\ell \ell_3 \ell_4}^{m m_3 m_4} \notag \\ & \times \avg{\Delta^L_{\ell_1m_1}(k)\Delta_{\ell_3m_3}^*(k)}
\end{align}
The window function therefore couples different multipoles \citep{Hiv02,HT95,Pratten13,Els17} and different wavenumber \citep{Liu16}. Further simplifications are possible,
\begin{align}
&\sum_m\avg{\tilde{\delta}_{\ell m}^L(k)\delta_{\ell m}^L(k)}
= \sum_{\ell_1 \ell_2} \frac{(2\ell+1)(2\ell_1+1)}{4\pi} \tj{\ell}{\ell_1}{\ell_2}{0}{0}{0}^2\notag \\
&C^W_{\ell_2}\frac{2}{\pi}\int \mathrm{d}s_1\,s_1^2\,\mathrm{d}s_2\,s_2^2\,C_{\ell_1}(s_1,s_2)j_{L}(k s_1)j_{\ell_1}(k s_2) \phi(s_1)\phi(s_2)
\end{align}
where $ C^W_{\ell_2} = \sum_{m_2} |W_{\ell_2,m_2}|^2$ is the power spectrum of the angular mask.

\section{Conclusions}
\label{sec:conclusions}

Since the observed redshifts of cosmological objects contain a component of their line-of-sight velocity, due to the Doppler effect, their clustering selects a preferred origin.  This breaks the assumption of statistical homogeneity which is frequently invoked in analyses of redshift survey data.  In the limit that all of the objects are close together on the sky translational symmetry is restored, so the violations are often referred to as ``wide angle effects''.  In this paper we have studied the impact of such effects on the estimation of the correlation function and power spectrum, including approximations which are often made in the analysis of redshift surveys.

For the 2-point functions in configuration and Fourier space we have presented new expressions for the impact of wide angle effects through to order $\mathcal{O}(\theta^2)$, which is also quadratic order in the ratio of the pair separation to the distance from the observer or the product of the wavenumber and the observer distance. The often neglected wide angle contribution from the galaxy selection function are also discussed. We find that for a homogeneous galaxy sample the physical wide angle effects are partially canceled by the ones introduced by the selection function. For more general cases however this is unlikely to happen.
We found that for the low multipole moments which dominate the signal, wide angle effects are generally small.  Even so, they can be accounted for in the modeling in a simple manner as long as the perturbations are linear at large scales.

Some of the new terms, although small, are out of phase with the standard multipoles in the plane-parallel limit. We quantify this difference and find 1\% to 0.1\% effects at the BAO scale, with lower redshift being more problematic.
This feature of the wide angle terms does not automatically translate into a bias in distance estimate of the BAO, but it should be kept in mind in the quest for sub percent BAO measurements.

We find that the loss of translational invariance changes the way angular and radial masks affect the measurements in Fourier space. This problem cannot be written anymore as a simple convolution, as often assumed in data analysis, and it is actually much more similar to the CMB case, where different multipoles of the underlying power spectrum and the windows are coupled.

The third kind of wide angle effects we studied are the ones introduced by an approximate LOS choice in power spectrum estimators. Our results show these biases are much bigger than the physical wide angle corrections, although still smaller than the cosmic variance error for a typical survey observing a large fraction of the sky. 

We would like to stress again that the importance of having an analytical understanding of these three wide angle effects is that we can now forward model them into the analysis for the cases where they could lead to potential systematic biases. Since they manifest on very large scales, our linear theory calculation is fairly accurate.

In the second part of this work we focused on the (spherical) Fourier-Bessel expansion of cosmological fields, which naturally respects the radial symmetry of the problem and isolates the redshift and `sky' directions.  We compare this formalism to the more standard Fourier analysis, and comment on the wide-angle effects and the impact of survey geometry. We presented new estimators, for both the power spectrum and the bispectrum, constructed with sFB coefficients, that are nicely related to the analytical models of galaxy clustering, in this way solving one of the major issues of spherical analysis.

Finally we pointed out how systematics in the purely angular domain can be much more robustly isolated in a spherical analysis, as the separation of scales is done at the level of the fields and not of the estimated correlation function. Our formalism is general in the sense it does not depend on the particular effect one wants to remove, and it applies as well to any higher order statistics of the galaxy field and its cross correlations with other probes.

\vspace{0.2in}
We would like to thank Nikhil Padmanabhan 
for useful discussions and initial collaboration during the early stage of this work. EC would like to thank Uros Seljak, Yin Li and Zack Slepian for useful discussion on power spectrum estimators.
M.W.~is supported by the  U.S. Department of Energy and by NSF grant number 1713791.
This work made extensive use of the NASA Astrophysics Data System and of the {\tt astro-ph} preprint archive at {\tt arXiv.org}. 

\appendix

\section{Geometry}
\label{app:geometry}

In this appendix we give some useful results for the triangle shown in Fig.~\ref{fig:triangle}.
Given $\vec{s}_1$ and $\vec{s}_2$ with $\vec{s}\equiv\vec{s}_1-\vec{s}_2$
we have defined the line of sight parallel to the angle bisector
\begin{equation}
  \vec{d} = \frac{s_1s_2}{s_1+s_2}\left(\hat{s}_1+\hat{s}_2\right)
\end{equation}
with (squared) length
\begin{align}
\label{eq:lengths}
  d^2 &= s_1s_2\left[1-\frac{(\vb{s}_1-\vb{s}_2)^2}{(s_1+s_2)^2}\right] \\
  &= \frac{4s_1^2s_2^2}{(s_1+s_2)^2}\cos^2\frac{\theta}{2}
  \quad .
\end{align}
By considering $s_1^2=|(1-t)\vec{s}+\vec{d}|^2$ and $s_2^2=|\vec{s}t-\vec{d}|^2$
and eliminating the $\hat{s}\cdot\hat{d}$ terms we have Stewart's theorem
in the form:
\begin{equation}
  s_1^2 s t + s_2^2 s (1-t)
  = s \left[d^2+s^2 t(1-t)\right]
\end{equation}
In combination with $t s_1=(1-t)s_2$ this gives
\begin{equation}
  d^2 + s^2 t(1-t) = s_1s_2 = s_1^2\frac{t}{1-t}
\end{equation}
which can be solved to yield
\begin{equation}
  1-t = \frac{-1+\mu x+\sqrt{1+\mu^2 x^2}}{2\mu x}
  \simeq \frac{1}{2} + \frac{\mu x}{4} - \frac{\mu^3 x^3}{16} + \cdots
\end{equation}
for $x\equiv s/d$ and $\mu=\cos\phi$.  It is helpful to expand
\begin{eqnarray}
  s_1 &\simeq& d\left[1+\frac{\mu x}{2}+\frac{(1+\mu^2)x^2}{8} + \cdots\right]\\
  s_2 &\simeq& d\left[1-\frac{\mu x}{2}+\frac{(1+\mu^2)x^2}{8} + \cdots\right]
\end{eqnarray}
for small $x$.
Starting from $\hat{d}\cdot\hat{s}_1$ we find
\begin{eqnarray}
  \cos\frac{\theta}{2}
  &=& \frac{1+x\mu(1-t)}{\sqrt{1+2x(1-t)\mu+x^2(1-t)^2}} \\
  &\simeq& 1 - \frac{(1-\mu^2)x^2}{8} + \cdots \\
  \sin\frac{\theta}{2}
  &\simeq& \frac{x}{2}\sqrt{1-\mu^2} + \cdots
\end{eqnarray}
which implies
\begin{equation}
  \mathcal{L}_\ell\left(\cos\frac{\theta}{2}\right) \simeq
  1 + x^2\frac{\ell(\ell+1)}{16}\left[\mu^2-1\right] + \cdots
\end{equation}
as stated in the main text.
It is also useful to consider $\mathcal{L}_\ell(\cos\theta)$ in the limit of small $\theta$ but possibly large $\ell$.  This can be rewritten as
\begin{equation}
  \mathcal{L}_\ell(\cos\theta)\simeq J_0(\ell\tilde{\omega})
  \simeq J_0(\ell\,x\sqrt{1-\mu^2})
\end{equation}
with $\tilde{\omega}\equiv 2\sin(\theta/2)\ll 1$ and $J_0$ the cylindrical Bessel function of order $0$.

The other common choice for the line of sight is the mid-point
\begin{equation}
  \vec{d}_{m} \equiv \frac{1}{2}\left(\vec{s}_1+\vec{s}_2\right)
  = \vec{d} + \left(\frac{1}{2}-t\right)\vec{s}
  \quad .
\end{equation}
This is equal to the $\vec{d}$ above in the limit $x\to 0$, but there are differences outside of this limit.  For small $x$ the length
\begin{equation}
  d_{m} = d\left[ 1+\frac{\mu^2 x^2}{4}+\cdots\right]
\end{equation}
while the angle to the line of sight becomes
\begin{equation}
  \mu_{m} = \hat{s}\cdot\hat{d}_{m}
  = \mu\left[ 1 + \frac{x^2}{4}(1-\mu^2)+\cdots\right]
\end{equation}
These relations, and their obvious inverses, can be used to express $\xi(s,d_{m},\mu_{m})$ in terms of $\xi(s,d,\mu)$ for small $x$.

Finally we can define the triangle in terms of $\vec{s}_1$ and $\vec{s}$, with $\mu_1=\hat{s}\cdot\hat{s}_1$ and expansion parameter $x_1=s/s_1\ll 1$.  The relation to the expansion in the text is then
\begin{align}
  \hat{s}_1\cdot\hat{s}_2 &\simeq 1 - \frac{x_1^2}{2}(1-\mu_1^2) + \cdots \\
  \mu &\simeq \mu_1 - \frac{x_1}{2}(1-\mu_1^2) + \cdots \\
  x   &\simeq x_1\left(1 + \frac{\mu_1x_1}{2}\right) + \cdots
\end{align}


\section{Useful identities}
\label{app:identities}

In this appendix we collect some identities which are useful in deriving the formulae in the main text.  The conversion from Fourier transforms to multipoles is accomplished using the Rayleigh expansion of a plane wave:
\begin{equation}
  e^{i\vec{k}\cdot\vec{r}} = \sum_\ell i^\ell(2\ell+1)j_\ell(kr)
  \mathcal{L}_\ell(\hat{k}\cdot\hat{r})
\end{equation}
The (spherical) Bessel functions satisfy a completeness relation
\begin{equation}
  \int s^2\,ds\ j_\ell(ks)j_\ell(k's) = \frac{\pi}{2kk'}
  \delta^{(D)}(k-k')
\end{equation}
The spherical harmonics obey the addition theorem
\begin{equation}
  \mathcal{L}_\ell(\hat{r}_1\cdot\hat{r}_2) = \frac{4\pi}{2\ell+1}
  \sum_m Y_{\ell m}(\hat{r}_1)Y_{\ell m}^\star(\hat{r}_2)
\end{equation}
while the solid harmonics
\begin{equation}
  R_\ell^m(\mathbf{r}) \equiv \sqrt{\frac{4\pi}{2\ell+1}}
  \ r^\ell\ Y_\ell^m(\hat{r})
\end{equation}
obey an analogous addition theorem
\begin{equation}
  R_\ell^m\left(\mathbf{x}+\mathbf{y}\right) = \sum_{\lambda=0}^\ell
  \sum_{\mu=-\lambda}^\lambda
  R_\lambda^\mu(\mathbf{x}) R_{\ell-\lambda}^{m-\mu}(\mathbf{y})
  \binom{\ell+m}{\lambda+\mu}^{1/2}
  \binom{\ell-m}{\lambda-\mu}^{1/2}
\end{equation}
The spherical harmonic addition theorem, and the orthogonality of the $Y_{\ell m}$, can be used to prove
\begin{equation}
  \int d\hat{x}\ \mathcal{L}_L(\hat{k}\cdot\hat{x})
  \mathcal{L}_\ell(\hat{s}\cdot\hat{x}) =
  \frac{4\pi}{2L+1}\delta_{L\ell}\mathcal{L}_{L}(\hat{k}\cdot\hat{s})
\end{equation}
In combination with the Rayleigh expansion this implies
\begin{equation}
\int\frac{d\Omega_k}{4\pi} e^{i\vec{k}\cdot\vec{s}}
\mathcal{L}_L(\hat{k}\cdot\hat{d}) = i^L j_L(ks)
\mathcal{L}_L(\hat{s}\cdot\hat{d})
\end{equation}
Finally, we note that the $3j$ symbols obey
\begin{equation}
\label{eq:threejI}
  \sum_{M,m_1} \tj{\ell_1}{\ell_2}{L}{m_1}{m_2}{M}
  \tj{\ell_1}{\ell}{L}{m_1}{m}{M}
  = \frac{1}{2\ell+1} \delta_{\ell\ell_2}\delta_{mm_2}
\end{equation}

\section{Derivation of linear theory result}
\label{app:derivation_sketch}

We take the expressions for the linear theory correlation function from \citet{Sza98}, who give the expression for an arbitrary triangle configuration.  It is useful to briefly recap how that derivation proceeds, so we give an outline of some of the steps below for completeness.  To keep the derivation as short as possible we only show some terms and in particular we omit the $\alpha$ terms.  The other terms follow a similar pattern and can be found in \citet{Sza98} if desired.

Recall the redshift-space density in linear theory is \citep{Kai87}
\begin{equation}
  \delta^{(s)}(\vb{s}) = \int\frac{d^3k}{(2\pi)^3}
  e^{i\vb{k}\cdot\vb{s}}
  \left(1+\beta[\hat{k}\cdot\hat{s}]^2\right)\delta^{(r)}(\vb{k})
\end{equation}
If we define
\begin{equation}
  \delta_\ell \equiv \int \frac{d^3k}{(2\pi)^3}
  \mathcal{L}_\ell(\hat{k}\cdot\hat{s}) e^{i\vb{k}\cdot\vb{s}}\delta(\vb{k})
\end{equation}
then using $\mu^2=(2/3)\mathcal{L}_2(\mu)+(1/3)\mathcal{L}_0(\mu)$ we have
\begin{equation}
  \delta^{(s)}(\vb{s}) = \left(1+\frac{\beta}{3}\right)\delta_0 +
  \frac{2\beta}{3}\delta_2
\end{equation}
The correlation function is thus
\begin{align}
  \xi(\vb{s}_1,\vb{s}_2) &= \left(1+\frac{\beta}{3}\right)^2
  \left\langle\delta_0\delta_0\right\rangle +
  \frac{4}{9}\beta^2\left\langle\delta_2\delta_2\right\rangle
  \notag \\ 
  &+ \frac{2\beta}{3}\left(1+\frac{\beta}{3}\right)
  \left\langle\delta_0\delta_2+\delta_2\delta_0\right\rangle
\end{align}
To evaluate the expectation values we expand $\mathcal{L}$ and $\exp[i\vb{k}\cdot\vb{s}]$ in spherical harmonics and integrate over $d\Omega_k$.  For example
\begin{equation}
  \left\langle \delta_0^2\right\rangle = \int\frac{k^2\,dk}{2\pi^2}P(k)j_0(ks)
\end{equation}
as expected while
\begin{align}
\left\langle\delta_0\delta_2+\delta_2\delta_0\right\rangle =& -\left[\mathcal{L}_2(\hat{s}\cdot\hat{s}_1)+\mathcal{L}_2(\hat{s}\cdot\hat{s}_2)\right] \notag \\
 & \int\frac{k^2\,dk}{2\pi^2}P(k)j_2(ks) \\
 =& -\left[2\mathcal{L}_2(\mu)\cos\theta+\frac{1}{2}\left(1-\cos\theta\right)\right] \notag \\
 & \int\frac{k^2\,dk}{2\pi^2}P(k)j_2(ks)
\end{align}
and
\begin{align}
  \left\langle \delta_2^2\right\rangle &= \int\frac{k^2\,dk}{2\pi^2}P(k)\sum_{L} i^Lj_L(ks)\ \left(\frac{4\pi}{5}\right)^2 \notag \\
  &\times \sum_{M,m_1,m_2}
  \mathcal{G}_{L22}^{Mm_1m_2}Y_{LM}^\star(\hat{s})
  Y_{2m_1}^\star(\hat{s}_1)Y_{2m_2}^\star(\hat{s}_2)
\end{align}
To evaluate the last line, set $\hat{d}$ to be the $\hat{z}$-axis and orient the triangle (Fig.~\ref{fig:triangle}) to lie in the $x-z$ plane (so all of the polar angles are zero or $\pi$).  Note that $\hat{s}_1$ and $\hat{s}_2$ are both at angle $\theta/2$ to $\hat{d}$ while $\hat{s}$ is at angle $\pi-\phi$.    Only $0\le L\le 4$ are non-zero and using the explicit forms of the $Y_{\ell m}$ then gives
\begin{equation}
  \left\langle \delta_2^2\right\rangle \ni  \frac{\mathcal{L}_2(\cos\theta)}{5}\int\frac{k^2\,dk}{2\pi^2}P(k)j_0(ks)
\end{equation}
for the $L=0$ contribution
\begin{align}
  \left\langle \delta_2^2\right\rangle & \ni  \frac{1}{28}\left[1-3\cos(2\theta)
  \right. \notag \\
  & \left. -3\cos(2\phi-\theta)-3\cos(2\phi+\theta)\right] \notag \\
  & \times\int\frac{k^2\,dk}{2\pi^2}P(k)j_2(ks)
\end{align}
for the $L=2$ contribution and
\begin{align}
  \left\langle \delta_2^2\right\rangle & \ni  \frac{9}{1120}\left[6+35\cos(4\phi)+3\cos(2\theta)+
  \right. \notag \\
  & \left. 10\cos(2\phi-\theta)+10\cos(2\phi+\theta)\right] \notag \\
  & \times\int\frac{k^2\,dk}{2\pi^2}P(k)j_4(ks)
\label{eq:L4contribution}
\end{align}
for $L=4$.  Note that in the limit $\theta\to 0$
\begin{equation}
  \left(1+\frac{\beta}{3}\right)^2 + \frac{4\beta^2}{45}\mathcal{L}_2(\cos\theta) \to
  1 + \frac{2}{3}\beta + \frac{1}{5}\beta^2
\end{equation}
and the $L=4$ part of
\begin{equation}
  \frac{4\beta^2}{9}\left\langle \delta_2^2\right\rangle
  \to \frac{8\beta^2}{35}\mathcal{L}_4(\mu)
  \int\frac{k^2\,dk}{2\pi^2}P(k)j_4(ks)
\end{equation}
as desired.  The other terms follow a similar pattern, and the results\footnote{Note that Eq.~(15) of \citet{Sza98} contains a typographical error.  The $4/15$ should be $8/15$.} can be found in \citep[][beware that their $\theta$ is half ours]{Sza98}.  It is easy to show that the corrections to the plane-parallel limit start at $\mathcal{O}(\theta^2)$.

\section{Linear theory redshift-space angular multipoles}
\label{app:lin_expression}

In the main text we showed that the redshift-space angular multipoles in linear theory could be written 
\begin{align}
\label{eq:Clk}
C_\ell(k_1,k_2) = & \int \mathrm{d} k \,k^2 P(k)  \int \mathrm{d}s_1\,s_1^2\,\int \mathrm{d}s_2\,s_2^2
\,j_{\ell}(k_1 s_1)j_{\ell}(k_2 s_2) \notag \\
& [j_{\ell}(ks_1)-f j_{\ell}''(ks_1)]
  [j_{\ell}(ks_2)-f j_{\ell}''(ks_2)]
\end{align}
This expression is different than the one originally appearing in \citep{HT95,Fisher94}, but it rather follows from Hankel-transforming the MAPS of \citep{Dat07}.\eq{eq:Clk} contains two kind of integrals over the product of spherical Bessel function. The first one is the orthogonality  
relation
\begin{equation}
\label{eq:jd}
\int \mathrm{d}s\,s^2 j_\ell (k s) j_\ell( k_1 s) = \frac{\pi}{2}\frac{\delta^{(D)}(k-k_1)}{k^2}
\end{equation}
The second one involves the second derivative of the Bessel function
\begin{equation}
\int \mathrm{d}s\,s^2 j_\ell (k s) j_\ell''( k_1 s) 
\end{equation}
and can be analytically computed using Eq.~(1) and Eq.~(8) on page 405 of \cite{Watson}
\begin{align}
\label{eq:jj}
&\int \mathrm{d}s\,s^2 j_\ell (k s) j_\ell''( k_1 s)= \notag \\
&\frac{\pi}{2}\left[\theta(k_1-k)\frac{(\ell+1)(\ell+2)}{2 \ell+1}\frac{k^{\ell+1/2}}{k_1^{\ell+5/2}}+\theta(k-k_1)\frac{\ell(\ell-1)}{2\ell+1}\frac{k_1^{\ell-3/2}}{k^{\ell+1/2}}\right]
\end{align}
\eqs{eq:jd}{eq:jj} can then be plugged into \eq{eq:Clk} to perform the remaining integral in $k$.

\section{Estimators at $\mathcal{O}(\theta^2)$}

By definition there is no way to make the Yamamoto estimator fully separable. However it is possible, order by order in the small angle between the pair of objects, to construct a separable estimator. In configuration space an estimator accurate to $\mathcal{O}(\theta^2)$ has been presented in \cite{Slepian15}.
We can simplify the expression for $P_L(k)$ by using the addition theorem for solid harmonics (see Appendix \ref{app:identities})
\begin{align}
  P_L^Y(k) &=
  \int \mathrm{d}\Omega_\mathbf{k}
  \,\mathrm{d}^3s_1\,\mathrm{d}^3s_2
  \ \delta(\vb{s}_1) \delta(\vb{s}_2) e^{-i\vb{k}\cdot \vb{s}}
  \notag \\ & \sum_M Y_{LM}^\star(\hat{k})Y_{LM}(\hat{d}) \\
   &= \int \mathrm{d}\Omega_\mathbf{k}
  \,\mathrm{d}^3s_1\,\mathrm{d}^3s_2
  \ \delta(\vb{s}_1) \delta(\vb{s}_2) e^{-i\vb{k}\cdot \vb{s}} \notag \\
  & \left(\frac{2L+1}{4\pi}\right)^{1/2} \sum_M Y_{LM}^\star(\hat{k})
  R_{L}^M(\vb{d})d^{-L} \\
  &  = \sum_{\ell m M} \int \mathrm{d}\Omega_\mathbf{k} Y_{LM}^\star(\hat{k})
  \,\mathrm{d}^3s_1\,\mathrm{d}^3s_2
  \ \delta(\vb{s}_1) \delta(\vb{s}_2) e^{-i\vb{k}\cdot\vb{s}} \notag \\ 
  & \left(\frac{2L+1}{4\pi}\right)^{1/2} \left(\frac{4\pi}{2\ell+1}\right)^{1/2}
  \left(\frac{4\pi}{2(L-\ell)+1}\right)^{1/2} \notag \\
  & \binom{L+M}{\ell+m}^{1/2}
  \binom{L-M}{\ell-m}^{1/2}\left(\frac{ts_1}{d}\right)^L \notag \\
  & Y_{L-\ell,M-m}(\hat{s}_1)Y_{\ell m}(\hat{s}_2)
\end{align}
where we have used $\vb{d}=t\vb{s}_1+(1-t)\vb{s}_2$ and $ts_1=(1-t)s_2=s_1s_2/(s_1+s_2)$ in the last line.  The spherical harmonics in $\hat{s}_1$ and $\hat{s}_2$ are now separated, but $t$ and $d$ are still functions of the angle between the direction of the two galaxies. With the help of \eq{eq:lengths} we can write 
\begin{align}
\left(\frac{ts_1}{d}\right)^L = \left[2\cos\frac{\theta}{2}\right]^{-L} &= 2^{-L/2}(1+\cos \theta)^{-L/2}\notag\\ &\simeq 2^{-L}\left[\left(1+\frac{L}{4}\right)-\frac{L}{4}\cos \theta\right]
\end{align}
therefore obtaining a power spectrum estimator as the sum of a $\mathcal{O}(\theta^0)$ and a $\mathcal{O}(\theta^2)$ term
\begin{align}
P_L^Y(k) &\simeq  \sum_{\ell m M} \int \mathrm{d}\Omega_\mathbf{k} Y_{LM}^\star(\hat{k})
  \,\mathrm{d}^3s_1\,\mathrm{d}^3s_2
  \ \delta(\vb{s}_1) \delta(\vb{s}_2) e^{-i\vb{k}\cdot\vb{s}} \notag \\ 
  & \left(\frac{2L+1}{4\pi}\right)^{1/2} \left(\frac{4\pi}{2\ell+1}\right)^{1/2}
  \left(\frac{4\pi}{2(L-\ell)+1}\right)^{1/2} \notag \\
  & \binom{L+M}{\ell+m}^{1/2}
  \binom{L-M}{\ell-m}^{1/2} Y_{L-\ell,M-m}(\hat{s}_1)Y_{\ell m}(\hat{s}_2) \notag \\
  & 2^{-L}\left[\left(1+\frac{L}{4}\right)-\frac{L}{4}\cos \theta\right] \notag \\
  & \equiv P_L^{Y,(0)}(k) +  P_L^{Y,(2)}(k)
\label{eq:Pktheta}
\end{align}
The zero-th order piece is trivially separable as sum of FFT/sFB transforms, and similarly for $P_L^{Y,(2)}(k)$ we have
\begin{align}
P_L^{Y,(2)}(k) &= -\frac{4 \pi/3 L}{2^{L+2}}\sum_{M_1}\sum_{\ell m M} \sum_{\ell_1 m_1}\sum_{\ell_2 m_2} \mathcal{G}_{1, L-\ell, \ell_1}^{M_1, M-m, m_1}\mathcal{G}_{1, \ell, \ell_2}^{M_1, m ,m_2}\notag\\
& \int \mathrm{d}\Omega_\mathbf{k} Y_{LM}^\star(\hat{k})
  \,\mathrm{d}^3s_1\,\mathrm{d}^3s_2
  \ \delta(\vb{s}_1) \delta(\vb{s}_2) e^{-i\vb{k}\cdot\vb{s}} \notag \\ 
  & \left(\frac{2L+1}{4\pi}\right)^{1/2} \left(\frac{4\pi}{2\ell+1}\right)^{1/2}
  \left(\frac{4\pi}{2(L-\ell)+1}\right)^{1/2} \notag \\
  & \binom{L+M}{\ell+m}^{1/2}
  \binom{L-M}{\ell-m}^{1/2} Y_{\ell_1m_1}(\hat{s}_1)Y^\star_{\ell_2 m_2}(\hat{s}_2)
\end{align}
which contains only a finite number of terms.
Notice that $P_L^{Y,(0)}$ does not reduce to the standard FFT estimator in \eq{eq:PkFFT}, showing once again how the assumption $\hat{d}\simeq\hat{s}_1\simeq\hat{s}_2$ in \eq{eq:PkFFT} is not the result of a well defined series expansion but rather of an approximation.
In \eq{eq:Pktheta} we have split the estimator in two  terms to highlight the different contribution in $\theta$, but for practical reasons, \eg numerical convergence of the integrals, it is highly recommended to sum the two in the integrand prior performing the FFT/sFB transform. 
The estimator we just described could be useful to compare theoretical models with measurements on the largest scales, as the only wide angle contributions  at this order will be the physical ones we described in Section \ref{sec:linear_theory}.
\section{The Bispectrum estimator}
\label{sec:Bisp}

\citet{Sco15} demonstrated how to construct FFT estimators for the multipoles of the galaxy bispectrum with respect to the largest size of the triangle formed by the three wavenumbers. Not all of the bispectrum information is contained in these multipoles, but a Fisher analysis in \citet{Gagrani} showed that they provide most of the constraining power. The definition in \citet{Sco15} is
\begin{align}
\hat{B}_L (k_1,k_2,k_3) =& \frac{2L+1}{N}\int \frac{\mathrm{d}^3 q_1}{(2\pi)^3}\,\int \frac{\mathrm{d}^3 q_2}{(2\pi)^3}\,\int \frac{\mathrm{d}^3 q_3}{(2\pi)^3} \notag \\&(2\pi)^3\delta_D^{(3)}(\mathbf{q}_1+\mathbf{q}_2+\mathbf{q}_3)\times \notag \\
& \int \mathrm{d}^3 s_1\,\mathrm{d}^3 s_2\,\mathrm{d}^3 s_3 \,\delta(\mathbf{s_1})\delta(\mathbf{s_2})\delta(\mathbf{s_3}) \notag \\ &\times \mathcal{L}_L(\hat{q}_1\cdot\hat{s}_1)e^{-i \mathbf{q}_1\cdot\mathbf{s}_1-i \mathbf{q}_2\cdot \mathbf{s}_2-i\mathbf{q}_3\cdot\mathbf{s}_3}
\end{align}
where $N$ is a normalization factor that depends on the particular configuration, and the integrals over the $q_i$ are evaluated around a thin shell $q_i = k_i +\delta k_i$.

It turns out that this can be computed relatively simply within the Fourier-Bessel formalism.  To make connection with the sFB expansion, let us rewrite the $\delta_D^{(3)}$ as
\begin{equation}
(2\pi)^3\delta_D^{(3)}(\mathbf{q}_1+\mathbf{q}_2+\mathbf{q}_3) = \int \mathrm{d}^3 s \ e^{i(\mathbf{q}_1+\mathbf{q}_2+\mathbf{q}_3)\cdot \vb{s}}
\end{equation}
and then perform all the angular integrals over the wavenumbers and use the $3j$ identity \eq{eq:threejI}
\begin{align}
\label{eq:Bisp}
\hat{B}_L(k_1,k_2,k_3) \propto &\sum_{\substack{\ell_1,\ell_2,\ell_3\\m_1,m_2,m_3}} \sum_{\ell<|\ell_1-L|}^{\ell_1+L} i^{\ell_1-\ell}\left(\frac{\pi}{2}\right)^{3/2} \notag \\
&(2L+1)(2l+1) \mathcal{G}_{\ell_1\ell_2\ell_3}^{m_1m_2m_3}
\tj{\ell}{\ell_1}{L}{0}{0}{0}^2 \notag \\ &
\int \mathrm{d}s\,s^2\,j_{\ell_1}(k_1 s)j_{\ell_2}(k_2 s)j_{\ell_3}(k_3 s)\notag\\
&\delta_{\ell_1m_1}^{\ell}(k_1)\delta_{\ell_2m_2}(k_2)\delta_{\ell_3m_3}(k_3)
\end{align}
The integral over three spherical Bessel functions can be computed analytically, following \citet{Mehrem+91}, as
\begin{align}
&\tj{\ell_1}{\ell_2}{\ell_3}{0}{0}{0} \int \mathrm{d}s\,s^2\,j_{\ell_1}(k_1 s)j_{\ell_2}(k_2 s)j_{\ell_3}(k_3 s) =\notag \\ 
&\frac{\pi}{4k_1k_2k_3}i^{\ell_1+\ell_2-\ell_3}(2\ell_3+1)^{1/2}\left(\frac{k_1}{k_3}\right)^{\ell_3} \notag \\ & \sum_{\ell_4=0}^{l_3}\binom{2 \ell_3}{2\ell_4}^{1/2} \left(\frac{k_2}{k_1}\right)^{\ell_4}
\sum_{\ell_5} (2\ell_5+1) \tj{\ell_1}{\ell_3-\ell_4}{\ell_5}{0}{0}{0}\notag \\ & \tj{\ell_2}{\ell_4}{\ell_5}{0}{0}{0}  \Gj{\ell_1}{\ell_2}{\ell_3}{\ell_4}{\ell_3-\ell_4}{\ell_5}\mathcal{L}_{\ell_5}(\hat{k}_1\cdot\hat{k}_2)
\end{align}
and it is different from zero only if the three wavenumbers live in a triangular configuration. This expression for the bispectrum may look cumbersome, but after the sFB coefficients have been computed, \eg for the power spectrum, estimating the bispectrum reduces to simple sums and multiplications.
Our expression automatically accounts for the fact the three modes live in a triangle, which in a Cartesian analysis usually requires additional FFTs. As already discussed in Sec\ref{sec:PksFB}, low $k_{||}$ modes can be discarded at the level of the field, therefore removing systematics in the plane of the sky does not present any extra work for the bispectrum.

\bibliographystyle{mnras}
\bibliography{}

\end{document}